\documentclass
[superscriptaddress,secnumarabic,amssymb,amsmath,nobibnotes,aps,prd,showkeys,showpacs,nofootinbib,onecolumn,notitlepage,12pt]{revtex4-2}%
\usepackage{setspace}
\usepackage{color}
\usepackage{amsmath}
\usepackage{amsfonts}
\usepackage{verbatim}
\usepackage{amssymb}
\usepackage{graphicx,bm}
\usepackage{graphicx}
\usepackage{amsmath}
\usepackage{amssymb}
\usepackage{amssymb}
\usepackage{graphicx,bm}
\usepackage{graphicx}
\usepackage[caption=false]{subfig}%
\providecommand{\U}[1]{\protect\rule{.1in}{.1in}}

\newcommand{\be}{\begin{equation}}
\newcommand{\ee}{\end{equation}}

\newcommand{\mincir}{\raise
-3.truept\hbox{\rlap{\hbox{$\sim$}}\raise4.truept\hbox{$<$}\ }}
\newcommand{\magcir}{\raise
-3.truept\hbox{\rlap{\hbox{$\sim$}}\raise4.truept\hbox{$>$}\ }}

\ifx\pdfoutput\relax\let\pdfoutput=\undefined\fi
\newcount\msipdfoutput
\ifx\pdfoutput\undefined\else
\ifcase\pdfoutput\else
\ifx\paperwidth\undefined\else
\ifdim\paperheight=0pt\relax\else\pdfpageheight\paperheight\fi
\ifdim\paperwidth=0pt\relax\else\pdfpagewidth\paperwidth\fi
\fi\fi\fi
\begin{document}
\title{Anisotropic Solutions in Symmetric Teleparallel $f\left(  Q\right)  $-theory:
Kantowski-Sachs and Bianchi III LRS
Cosmologies}
\author{N. Dimakis}
\email{nsdimakis@scu.edu.cn ; nsdimakis@gmail.com}
\affiliation{Center for Theoretical Physics, College of Physics, Sichuan University,
Chengdu 610065, China}
\author{M. Roumeliotis}
\email{microum@phys.uoa.gr}
\affiliation{Nuclear and Particle Physics section, Physics Department, University of
Athens, 15771 Athens, Greece}
\author{A. Paliathanasis}
\email{anpaliat@phys.uoa.gr}
\affiliation{Institute of Systems Science, Durban University of Technology, Durban 4000,
South Africa}
\affiliation{Departamento de Matem\'{a}ticas, Universidad Cat\'{o}lica del Norte, Avda.
Angamos 0610, Casilla 1280 Antofagasta, Chile}
\affiliation{Mathematical Physics and Computational Statistics Research Laboratory,
Department of Environment, Ionian University, Zakinthos 29100, Greece}
\author{T. Christodoulakis}
\email{tchris@phys.uoa.gr}
\affiliation{Nuclear and Particle Physics section, Physics Department, University of
Athens, 15771 Athens, Greece}

\begin{abstract}
We investigate the existence of anisotropic self-similar exact solutions in
symmetric teleparallel $f\left(  Q\right)  $-theory. For the background
geometry we consider the Kantowski-Sachs and the Locally Rotationally Symmetric Bianchi type III geometries. These
two anisotropic spacetimes are of special interest because in the limit of
isotropy they are related to the closed and open
Friedmann--Lema\^{\i}tre--Robertson--Walker cosmologies respectively. For each spacetime there exist two distinct families of flat, symmetric connections, which share the symmetries of the spacetime. We present the field equations, and from them, we determine the functional form of the $f\left(  Q\right)  $
Lagrangian which yields self-similar solutions. We initially consider the vacuum case and subsequently we introduce a matter source in terms of a perfect fluid. Last but not least, we report some
self-similar solutions corresponding to static spherically symmetric spacetimes.
\end{abstract}
\maketitle
\date{\today}

\section{ Introduction}

\label{sec1}

The affine connection, which describes the autoparallels, can be thought off as the fundamental object of a gravitational theory. From the symmetric, torsionless Levi-Civita connection, the Ricci scalar $R$ is derived, which consists the Lagrangian density of
General Relativity (GR) \cite{eingr}. From the curvature-less antisymmetric
Weitzenb{\"{o}}ck connection \cite{Weitzenb23}, which is related to the
existence of an unholonomic frame, the torsion scalar $T$ is constructed, which
leads to the Teleparallel Equivalence of General Relativity (TEGR)
\cite{Hayashi79}. Moreover, a torsion-free connection with zero Riemann tensor
and nonzero nonmetricity scalar $Q$, provides a gravitational Lagrangian
which results in field equations equivalent to those of General Relativity. This theory
is known as Symmetric Teleparallel General Relativity (STGR) \cite{mm1}.

The consideration of quantum-gravitational effects, described by a free quantum
field \cite{ha1}, introduces a quadratic term of the Ricci scalar in the
Einstein-Hilbert Action Integral \cite{ha2,ha3}. This modified gravitational
theory belongs to a more general family known as $f\left(  R\right)  $-theory
of gravity \cite{ha4,ha5,ha6}. The new degrees of freedom, which follow from
the modification of the Einstein-Hilbert Action Integral in $f\left(
R\right)  $-theory, can drive the dynamics so that to explain the late-time
acceleration phase attributed to dark energy \cite{fd1,fd2,fd3,fd4}; thus providing a geometric origin for the latter. These new degrees can also provide a
geometric description for the inflaton field \cite{ha2,fd5,fd6,fd7}, which
accounts for the early-time rapid expansion of the universe. In the context of
teleparallelism, the gravitational Action Integral has been modified in a
similar way as above, with the introduction of nonlinear components leading to
the family of teleparallel $f\left(  T\right)  $-theory of gravity \cite{st6}.
Indeed, the new geometrodynamical terms introduced in the field equations by
the nonlinear $f\left(  T\right)  $-theory, can explain the expanding history
of our universe \cite{st7,st8,st9,st10,st11,st12}. For a general review on various modified theories of gravity see \cite{Voik}.

Similarly, the symmetric teleparallel $f\left(  Q\right)  $-theory has been
introduced recently by cosmologists \cite{HeisfQ} in order to reconstruct the cosmological
history and explain the evolution of the cosmological parameters. It has been found that
cosmological models in $f\left(  Q\right)  $-theory can reproduce the $\Lambda
$CDM spacetime \cite{ww012,ww012a}, while comparison with cosmological
observations has shown that the $f\left(  Q\right)  $-theory is cosmologically
viable and can even challenge GR \cite{ww8,ww8b}. Moreover, it is
possible, in the context of $f\left(  Q\right)  $-theory, to have an effective fluid with a phantom behavior \cite{ww14,ww13}. Recently, in \cite{ww00}, self-similar cosmological solutions were derived in $f\left(  Q\right)  $-theory for a
Friedmann--Lema\^{\i}tre--Robertson--Walker (FLRW) background geometry with or
without spatial curvature. The existence of self-similar solutions is of
considerable interest, because they can describe asymptotic behaviors of more general solutions of the field
equations and also account for important eras of the cosmological history. For various other
studies of $f\left(  Q\right)  $-gravity in cosmology and in other types of spacetimes we refer the reader to
\cite{ww16,ww17,ww18,ww19,ww20,ww21,ww22,ww23,ww24,ww25,ww26,ww27,ww28,ww29,ww30,ww31,ww32,ww33,ww34,ww35,ww36,ww37} and references therein. It is important
to mention that while GR, TEGR and STGR are dynamically equivalent
theories of gravity, the same is not true in the case of $f-$theories \cite{DeFalco}. Indeed,
$f\left(  R\right)  $-theory is a fourth-order theory of gravity while
$f\left(  T\right)  $ and $f\left(  Q\right)  $ theories are of second-order.
In teleparallelism the selection of the proper vierbein is essential for the
theory. On the other hand, in symmetric teleparallel $f\left(  Q\right)
$-theory, the connection together with the metric are the fundamental fields.

In symmetric teleparallel theory the connection has a zero Riemann tensor,
i.e. $R_{\;\lambda\mu\nu}^{\kappa}=0$, which is referred as the flatness condition. As a result,
there exists a coordinate system, the coincident gauge, where all the
components of the connection vanish \cite{Eisenhart,Koi}. This means that, in
symmetric teleparallel theory, the inertial effects of the motion can be
separated from gravity. However, when the equations of motion are
considered after a partial gauge fixing at the level of the metric,
there may be an incompatibility between this choice of coordinates and the assumption of being in the coincidence gauge \cite{Zhao}. For the case of FLRW geometry, in $f\left(  Q\right)$-theory, there are three different connections compatible with zero spatial curvature, and just one for nonzero spatial curvature \cite{ww8c,ww8d,ww31}. For a similar analysis in the case of static spherical
symmetric spacetimes we refer the reader to \cite{nfq2}.

In this study, we are interested in the investigation of self-similar exact
solutions in $f\left(  Q\right)  $-theory for anisotropic cosmological models.
Specifically, for the Kantowski-Sachs (KS) and the Locally Rotationally Symmetric (LRS) Bianchi type III (BIII)
spacetimes. Both geometries admit a four-dimensional group of Killing
symmetries transitive on three-dimensional spacelike hypersurfaces. The groups consist of a translation symmetry and a
three-dimensional subgroup, the orbits of which are two-dimensional surfaces of
constant scalar curvature, positive for the Kantowski-Sachs geometry and
negative for the Bianchi III spacetime. The main characteristic of
the Kantowski-Sachs spacetime is that it can be seen as an extended
Schwarzschild manifold \cite{col1}. Some previous studies in the literature
on anisotropic spacetimes in symmetric teleparallel $f\left(  Q\right)
$-theory are presented in \cite{ww4,ww5,ww6,ww7}. In \cite{bb1,bb1b,bb1c} the
anisotropic Bianchi type I spacetime in the coincidence gauge in $f\left(
Q\right)  $-gravity was investigated, while the isotropic limit in such a
cosmology was studied in \cite{bb2}. The Bianchi type I geometry in the isotropic
limit reduces to the spatially flat FLRW geometry. On the other hand, the
Kantowski-Sachs and Bianchi III spacetimes in the isotropic asymptotic limit
are related to the closed and open FLRW models respectively. Anisotropies are
assumed to play an important role in the evolution of the very early
universe before the inflation \cite{kas4,kas5,kas6,kas7,kas8} and the existence of
exact solutions is of special interest for the viability of the gravitational
theory.

The structure of the paper is as follows: In Section \ref{sec2}, we briefly discuss the basic definitions of the
symmetric teleparallel $f\left(  Q\right)  $-theory. The self-similar metrics and the connections which share the symmetries of the spacetime are presented in Section \ref{sec3}. We show that there are two families of flat, symmetric connections compatible with the isometries of the spacetime, which produce three distinct cases if we additionally impose self-similarity. In Section \ref{sec4}, for each connection, we determine the explicit functional form for
the Lagrangian function, which is compatible with producing self-similar solutions. The presence of an
external fluid is also investigated. Furthermore, in Section \ref{sec5}, we demonstrate how the generic flat connections can be reconstructed from
the corresponding self-similar connections. In Section \ref{sec6}, we discuss the transformations connecting the generic connections to the reduced homothetic ones and how this affect the other fundamental field of the theory, the metric. Finally, in Section \ref{sec7}, we summarize our
results and draw our conclusions.

\section{Symmetric Teleparallel $f\left(  Q\right)  -$theory}

\label{sec2}

Consider a four dimensional manifold with a metric tensor $g_{\mu\nu}$ and the
covariant derivative $\nabla_{\mu}$ with the generic connection $\Gamma
_{\;\mu\nu}^{\kappa}$. We define the curvature tensor, $R_{\;\lambda\mu\nu}^{\kappa}$ as%
\begin{equation} \label{curvaturedef}
R_{\;\lambda\mu\nu}^{\kappa}\left(  \Gamma_{\;\mu\nu}^{\lambda}\right)
=\frac{\partial\Gamma_{\;\lambda\nu}^{\kappa}}{\partial x^{\mu}}%
-\frac{\partial\Gamma_{\;\lambda\mu}^{\kappa}}{\partial x^{\nu}}%
+\Gamma_{\;\lambda\nu}^{\sigma}\Gamma_{\;\mu\sigma}^{\kappa}-\Gamma
_{\;\lambda\mu}^{\sigma}\Gamma_{\;\mu\sigma}^{\kappa},
\end{equation}
the torsion, $T_{\mu\nu}^{\lambda}$,
\begin{equation} \label{torsiondef}
T_{\mu\nu}^{\lambda}\left(  \Gamma_{\;\mu\nu}^{\lambda}\right)  =\Gamma
_{\;\mu\nu}^{\lambda}-\Gamma_{\;\nu\mu}^{\lambda},
\end{equation}
and the nonmetricity, $Q_{\lambda\mu\nu}$,%
\begin{equation} \label{nonmdef}
Q_{\lambda\mu\nu}\left(  \Gamma_{\;\mu\nu}^{\lambda}\right)  = \nabla_{\lambda} g_{\mu\nu}=\frac{\partial
g_{\mu\nu}}{\partial x^{\lambda}}-\Gamma_{\;\lambda\mu}^{\sigma}g_{\sigma\nu
}-\Gamma_{\;\lambda\nu}^{\sigma}g_{\mu\sigma}.
\end{equation}

The Levi-Civita connection, $\hat{\Gamma}_{\;\mu\nu}^{\kappa}$, of General Relativity, is given by
\begin{equation} \label{LCcon}
  \hat{\Gamma}_{\;\mu\nu}^{\kappa} = \frac{1}{2} g^{\kappa\lambda} \left( \frac{\partial g_{\lambda\mu}}{\partial x^\nu} + \frac{\partial g_{\lambda\nu}}{\partial x^\mu} - \frac{\partial g_{\mu\nu}}{\partial x^\lambda} \right).
\end{equation}
If we set $\Gamma_{\;\mu\nu}^{\kappa}=\hat{\Gamma}_{\;\mu\nu}^{\kappa}$ inside \eqref{curvaturedef}, then we get a generally nonzero curvature tensor $\hat{R}_{\;\lambda\mu\nu}^{\kappa}= R_{\;\lambda\mu\nu}^{\kappa}\left(  \hat{\Gamma}_{\;\mu\nu}^{\lambda}\right)$, while
$T_{\mu\nu}^{\lambda}\left(  \hat{\Gamma}_{\;\mu\nu}^{\lambda}\right)
=0$ and $Q_{\lambda\mu\nu}\left(  \hat{\Gamma}_{\;\mu\nu}^{\lambda}\right)  =0$ are zero by construction.
The fundamental scalar of the theory is the Ricci scalar defined as
$\hat{R}=g^{\mu\nu}\hat{R}_{\mu\nu},$ where $\hat{R}_{\mu\nu}=g^{\kappa\nu}R_{\kappa\mu\nu\lambda}$.

For the Weitzenb{\"{o}}ck connection of teleparallelism,
$\tilde{\Gamma}_{\;\mu\nu}^{\kappa}$, it follows that
$R_{\;\lambda\mu\nu}^{\kappa}\left(  \tilde{\Gamma}_{\;\mu\nu}^{\lambda
}\right)=0  $ and $Q_{\lambda\mu\nu}\left(  \tilde{\Gamma}_{\;\mu\nu}^{\lambda
}\right)  =0$, while the torsion, $T_{\mu\nu}^{\lambda}\left(  \tilde{\Gamma}_{\;\mu\nu}^{\lambda}\right)$, is nonzero and given by \eqref{torsiondef}. The Lagrangian of TEGR consists of the torsion scalar, $T={S}_{\kappa}^{~~\mu\nu
}{T^{\kappa}}_{\mu\nu}$, where  ${S_{\beta}}^{\mu\nu}=\frac{1}%
{2}({K^{\mu\nu}}_{\beta}+\delta_{\beta}^{\mu}{T^{\theta\nu}}_{\theta}%
-\delta_{\beta}^{\nu}{T^{\theta\mu}}_{\theta})$ and $K_{~~~\beta}^{\mu\nu
}=-\frac{1}{2}({T^{\mu\nu}}_{\beta}-{T^{\nu\mu}}_{\beta}-{T_{\beta}}^{\mu\nu
})$ \cite{st6}.

In symmetric teleparallel theory of gravity the connection describes a flat, torsionless geometry, that
is, $R_{\;\lambda\mu\nu}^{\kappa}\left(  \Gamma_{\;\mu\nu}^{\lambda
}\right)  =0$ and $T_{\mu\nu}^{\lambda}\left( \Gamma_{\;\mu\nu
}^{\lambda}\right)  =0$, while the nonmetricity tensor is defined by \eqref{nonmdef}. The fundamental Lagrangian density of STGR, is given by the nonmetricity scalar, which is  \cite{mm1}%
\begin{equation}
Q=Q_{\lambda\mu\nu}P^{\lambda\mu\nu},\label{defQ}%
\end{equation}
where $P_{\;\mu\nu}^{\lambda}$ are the components of the nonmetricity
conjugate tensor
\begin{equation}
P_{\;\mu\nu}^{\lambda}=-\frac{1}{4}Q_{\;\mu\nu}^{\lambda}+\frac{1}{2}%
Q_{(\mu\phantom{\lambda}\nu)}^{\phantom{(\mu}\lambda\phantom{\nu)}}+\frac
{1}{4}\left(  Q^{\lambda}-\bar{Q}^{\lambda}\right)  g_{\mu\nu}-\frac{1}%
{4}\delta_{\;(\mu}^{\lambda}Q_{\nu)},\label{defP}%
\end{equation}
in which, $\delta_{\;\nu}^{\mu}$ is the Kronecker delta and where we introduce the contracted tensors, $Q_{\mu}=Q_{\mu\nu
}^{\phantom{\mu\nu}\nu}$ and $\bar{Q}_{\mu}=Q_{\phantom{\nu}\mu\nu}%
^{\nu\phantom{\mu}\phantom{\mu}}$.

\subsection{Field equations}

In symmetric teleparallel $f\left(  Q\right)  $-theory the gravitational
Action Integral is written as follows%

\begin{equation}
S=\frac{1}{2}\int d^{4}x\sqrt{-g}f(Q)+\int d^{4}x\sqrt{-g}\mathcal{L}_{M}
\label{action}%
\end{equation}
where $g=\mathrm{det}(g_{\mu\nu})$ is the determinant of the spacetime
metric and $\mathcal{L}_{M}$ is the Lagrangian function for the matter source.

The gravitational field equations\footnote{Formally, before the variation takes place, the Riemann curvature and the torsion tensor components should also be included in the action \eqref{action}, multiplied with Lagrange multipliers. The variation with respect to the multipliers then leads to the satisfaction of the flatness and torsionless conditions; for details see \cite{Heis,Hohmann}.}, derived by variation with respect to the metric, are \cite{Harko}
\begin{equation} \label{fieldm}
f^{\prime}(Q)G_{\mu\nu}+\frac{1}{2}g_{\mu\nu}\left(  f^{\prime}%
(Q)Q-f(Q)\right)  +2f^{\prime\prime}(Q)\left(  \nabla_{\lambda}Q\right)
P_{\;\mu\nu}^{\lambda}=T_{\mu\nu},
\end{equation}
where $f^{\prime}(Q)=\frac{df}{dQ}$ (throughout this work primes denote differentiation with respect to the arguments) and $T_{\mu\nu}$ is the
energy-momentum tensor for the matter source. The $G_{\mu\nu}=\hat{R}_{\mu\nu
}-\frac{1}{2}g_{\mu\nu}\hat{R}$ is the usual Einstein tensor, with
$\hat{R}_{\mu\nu}$ and $\hat{R}$ being the Riemannian Ricci tensor and
scalar respectively, constructed with the Levi-Civita connection \eqref{LCcon}. We observe
that for linear $f\left(  Q\right)  $ function, or for $Q=$const., the above field equations
reduce to those of General Relativity with a contribution from a cosmological constant. For this reason - and in order to study theories and dynamics beyond GR - in what follows, we shall restrict our attention in cases where $Q\neq$const., and where $f(Q)$ is a nonlinear function.

The equation of motion for the connection is derived as
\begin{equation}
\nabla_{\mu}\nabla_{\nu}\left(  \sqrt{-g}f^{\prime}(Q) P_{\phantom{\mu\nu}\sigma}^{\mu\nu}\right)  =0,\label{feq2}%
\end{equation}
while for a matter source minimally coupled to the metric the equations of
motion follow $T_{\phantom{\mu}\nu;\mu}^{\mu}=0$. The semicolon
\textquotedblleft$;$\textquotedblright\ here is used to denote the covariant
derivative with respect to the Levi-Civita connection.

In the following section, we proceed to investigate the particular cases of spacetimes we want to study and the connections which are compatible with their symmetries, so as to be used at the level of the equations of motion.

\section{Self-similar metrics and connections}

\label{sec3}

The Kantowski-Sachs and the LRS Bianchi type III geometries are described by a line element of the form \cite{MacCallum}
\begin{equation} \label{metric}
ds^{2}=-dt^{2}+a^{2}\left(  t\right)  dr^{2}+b^{2}\left(  t\right)
(d\theta^{2}+\Sigma_k(\theta)^2 d\phi^2) ,
\end{equation}
where $a\left(  t\right)$ and $b\left(  t\right)$ are the two scale
factors. The constant index, $k:=-\frac{\Sigma^{\prime\prime}_k(\theta)}{\Sigma_k(\theta)}$, is used to distinguish between the distinct models: For the Kantowski-Sachs we have $k=+1$, and $\Sigma_{+1}(\theta)=\sin\theta$, while for the LRS Bianchi type III model we use, $k=-1$ and $\Sigma_{-1}(\theta)=\sinh\theta$. We just mention here that the above line-element can also incorporate a LRS Bianchi type I spacetime if $k=0$ and $\Sigma_{0}(\theta)=\theta$. However, due to the fact that the type I model has considerably more admissible connections, we leave it outside the current work. The limit of isotropization\footnote{Due to the spatial geometry, only for the Bianchi type I case we can formally refer to full isotropy when $a=b$. For the Kantowski-Sachs and the Bianchi type III models, the term ``isotropization'' is used in the literature, as a reference to when the dynamics of the single remaining scale factor are, at some asymptotic limit, reminiscent of those of the FLRW spacetime of positive and negative spatial curvature respectively \cite{BBarrow}.} in the above models is attained for $\frac{a\left(  t\right)}{b\left(  t\right)  }\rightarrow1$. The admitted four Killing symmetries by the previous line element are%
\begin{equation} \label{killvecKS}
\begin{split}
& \xi_{1}=\partial_{\phi}~,~\xi_{2}=\cos\phi\partial_{\phi}- \frac{\Sigma^{\prime}_k(\theta)}{\Sigma_k(\theta)}\sin
\phi\partial_{\phi} \\
& \xi_{3}=\sin\phi\partial_{\phi}+\frac{\Sigma^{\prime}_k(\theta)}{\Sigma_k(\theta)} \cos\phi\partial_{\phi}\text{ and
}\xi_{4}=\partial_{r}.
\end{split}
\end{equation}

The requirement, for the geometries under study, to describe a self-similar
spacetime with a homothetic vector field $\xi_h$, i.e. satisfy the equation $\mathcal{L}_{\xi_h}g_{\mu\nu}=2 g_{\mu\nu}$, where $\mathcal{L}$ stands for the Lie derivative, constrains the scale factors
$a\left(  t\right)  $ and $b\left(  t\right)  $ so that, $a\left(  t\right)
=t^{1-\alpha}$ and $b\left(  t\right)  =\beta  t  $, where $\alpha$ and
$\beta$ are constants. Hence, the self-similar metric is
described by the line element
\begin{equation}
ds^{2}=-dt^{2}+t^{2(1-\alpha)}dr^{2}+\beta^{2}t^{2}(d\theta^{2}+\Sigma_k(\theta)^{2}
d\phi^2). \label{metricKS}%
\end{equation}
The admitted homothecy by the above line element (\ref{metricKS}) is%
\begin{equation}
\xi_{h}=t\partial_{t}+\alpha r\partial_{r}.
\end{equation}
We have written above the general solution of $\mathcal{L}_{\xi_h}g_{\mu\nu}=2 g_{\mu\nu}$, after clearing it out from unnecessary constants of integration, which either can be absorbed by diffeomorphisms or which are related to a trivial addition of Killing fields to $\xi_h$. The solution of the homothetic equation for this type of spacetimes in the coordinate system we use here can be found in \cite{KShom}, while conformal symmetries in general, for LRS spacetimes have been previously studied in \cite{Pantelis}.

Self-Similar solutions are known to play a prominent role in gravitational physics \cite{Carr,Carr2,Collins2}. This includes phenomena from astrophysics to cosmology. Self-similarity is used in many cases to model gravitational collapse, see \cite{LP1,LP2,sscol1,sscol2,sscol3,sscol4} for more details and examples on stellar collapse and self-similarity. Briefly, self-similar solutions describe naked singularity spacetimes which model the interior of collapsing stars, while the outside is being mapped to a vacuum solution, like Schwarzschild. Cosmological metrics like \eqref{metricKS} and also FLRW spacetimes can be used to describe such interior solutions. At the level of pure cosmology however, in which we are mostly preoccupied here, self-similarity is imposed in various additional contexts \cite{Eardley,Chao,Rosquist,Wainwright}. Self-similar solutions successfully reproduce various different eras in the evolution of the universe. In many cases, they can be used to approximate the asymptotic behaviour of more complicated solutions. This includes both the asymptotic limit towards the initial singularity, as well as the limit towards late times expansion \cite{Haager}. This has led to the postulation of the self-similarity hypothesis \cite{Carr}, namely that complicated gravitational systems can be, to some extent, approximated by self-similar configurations. For the above reasons, the study of self-similar solutions, even in the context of different gravitational theories, is still active \cite{ssnew1,ssnew2,ssnew3,ssnew4,ssnew5}.

In this work we are interested to see what $f(Q)$ theories can be compatible with self-similar gravitational solutions. In this type of theories, the connection plays an important role. It is true that the latter can be always set to zero by an appropriate coordinate transformation, trivializing the corresponding field equations. However, a part which is missed in various cases in the literature is that, when you assume a particular type of spacetime, e.g. some Bianchi type in cosmology, or some spherically symmetric spacetime, you already have spent part of the freedom of choosing a coordinate system. So, it may happen that the so called coincident gauge is incompatible with the coordinate system in which you have written the metric \cite{Zhao}. As a result it is not at all trivial to search for what type of connection can be used together with the class of metric you are assuming. Here, we adopt the strategy of requiring that the symmetries of the metric are shared also by the connection. Following a similar prescription to \cite{ww7,ww8c}, we thus demand that
\begin{equation} \label{Liedercon}
  \mathcal{L}_{\xi_i} \Gamma^{\lambda}_{\; \mu\nu} =  \xi_i^{\kappa} \frac{\partial \Gamma^{\lambda}_{\; \mu\nu} }{\partial x^\kappa } - \Gamma^{\kappa}_{\; \mu\nu} \frac{\partial \xi^{\lambda} }{\partial x^\kappa } + \Gamma^{\lambda}_{\; \kappa\nu} \frac{\partial \xi^{\kappa} }{\partial x^\mu } + \Gamma^{\lambda}_{\; \mu\kappa} \frac{\partial \xi^{\kappa} }{\partial x^\nu } + \frac{\partial^2 \xi^{\lambda}}{\partial x^\mu \partial x^\nu} =0,
\end{equation}
where $\xi_i$ is any of the four Killing vectors \eqref{killvecKS}.

We obtain two symmetric, and flat, connections satisfying relation \eqref{Liedercon}. Their nonzero components are seen below\footnote{It is understood that when we
write a particular component $\Gamma_{\mu\nu}^{\alpha}$, the $\Gamma_{\nu\mu}^{\alpha}$ has the exact same value.}. At most, two arbitrary functions of time $\gamma_1(t)$, $\gamma_2(t)$ and two constants $c_1$, $c_2$ are introduced.

\begin{itemize}

\item The connection $A$ contains $\gamma_{1}(t)$, $\gamma_{2}(t)$ and $c_{1}$
\begin{equation} \label{connectionA}
\begin{split}
&  \Gamma^{t}_{\;tt} = \gamma_{2}, \quad\Gamma^{r}_{\; tt} = \frac{1}{c_{1}}
\left(  \dot{\gamma}_{1} - \gamma_{1} \gamma_{2}+ \gamma_{1}^{2}\right)  ,
\quad\Gamma^{r}_{\; tr}= \Gamma^{\theta}_{\;t\theta} = \Gamma^{\phi}_{\;t\phi}
= \gamma_{1}, \quad\Gamma^{r}_{\; \theta\theta} = -\frac{k}{c_{1}},\\
&  \Gamma^{r}_{\; rr} =\Gamma^{\theta}_{\;r\theta}= \Gamma^{\phi}_{\;r\phi}=
c_{1}, \quad\Gamma^{r}_{\; \phi\phi} = -\frac{k }{c_{1}}\Sigma_k(\theta)^2,
\quad\Gamma^{\theta}_{\;\phi\phi} = -\Sigma_k(\theta)\Sigma^{\prime}_k(\theta), \\
& \Gamma^{\phi}_{\;\theta\phi} = \frac{\Sigma^{\prime}_k(\theta)}{\Sigma_k(\theta)}.
\end{split}
\end{equation}
The corresponding nonmetricity scalar is
\begin{equation} \label{nonmscA}
  Q = \frac{2 \left((2 \alpha -3) \beta ^2+k\right)}{\beta ^2 t^2}-\frac{3 (\alpha -3) \gamma_1}{t}+3 \dot{\gamma}_1
\end{equation}

\item The connection $B$ includes $\gamma_{1}(t)$, $\gamma_{2}(t)$, $c_{1}$ and $c_{2}$.

\begin{equation}%
\begin{split} \label{connectionB}
&  \Gamma^{t}_{\;tt} = -\frac{1}{\gamma_{2}}\left[  \dot{\gamma}_{2} +c_{1}
\gamma_{1} \left(  2-c_{2} \gamma_{1}\right)  + k \right]  , \quad\Gamma
^{t}_{\;tr} = c_{1} \left(  1- c_{2} \gamma_{1} \right)  , \quad\Gamma
^{t}_{\;rr} = c_{1} c_{2} \gamma_{2} ,\\
&\Gamma^{t}_{\;\theta\theta} = \gamma_{2},\quad \Gamma^{t}_{\;\phi\phi} = \gamma_{2} \Sigma_k(\theta)^2, \quad\Gamma^{r}_{\;
tt} = \frac{1}{\gamma_{2}^{2}} \left[  \gamma_{1} \left(  k+c_{1} \gamma_{1}
\right)  \left(  c_{2} \gamma_{1}-1\right)  -\gamma_{2} \dot{\gamma}_{1}
\right]  , \\
& \Gamma^{r}_{\; tr} = -\frac{c_{2} \gamma_{1}}{\gamma_{2}%
}\left(  k +c_{1} \gamma_{1}\right)  ,\quad \Gamma^{r}_{\; rr} = c_{1}+c_{2} k +c_{1}c_{2}\gamma_{1}, \quad\Gamma^{r}_{\;
\theta\theta} = \gamma_{1}, \\
& \Gamma^{r}_{\; \phi\phi} = \gamma_{1}\Sigma_k(\theta)^2, \quad \Gamma^{\theta}_{\;t\theta} = \Gamma^{\phi}_{\;t\phi} =
-\frac{k+c_{1} \gamma_{1}}{\gamma_{2}},\quad \Gamma^{\theta}_{\;r\theta}= \Gamma^{\phi}_{\;r\phi} = c_{1}, \\
& \Gamma^{\theta}_{\;\phi\phi} = -\Sigma_k(\theta)\Sigma^{\prime}_k(\theta), \quad \Gamma^{\phi
}_{\;\theta\phi} = \frac{\Sigma^{\prime}_k(\theta)}{\Sigma_k(\theta)}.
\end{split}
\end{equation}
In this case, the resulting nonmetricity scalar becomes
\begin{equation} \label{nonmscB}
  \begin{split}
    Q = & \dot{\gamma}_2 \left(\frac{(c_2 \gamma_1+2) (c_1 \gamma_1+k)}{\gamma_2^2}+c_1 c_2 t^{2 \alpha -2}+\frac{2}{\beta ^2 t^2}\right) - \frac{\dot{\gamma}_1 (2 c_1 c_2 \gamma_1+2 c_1+c_2 k)}{\gamma_2} \\
    & +\frac{\gamma_1 ((\alpha -3) (2 c_1+c_2 k))}{t \gamma_2}+\frac{\gamma_2 \left(-2 \alpha +(\alpha +1) \beta ^2 c_1 c_2 t^{2 \alpha }+2\right)}{\beta ^2 t^3} \\
    & +\frac{ c_1 c_2(\alpha -3) \gamma_1^2  }{t \gamma_2}+\frac{2}{t^2} \left(2 \alpha +\frac{k}{\beta ^2}+\frac{(\alpha -3) k t}{\gamma_2}-3\right).
  \end{split}
\end{equation}

\end{itemize}

Moreover, if we use $\xi_{h}$ to impose further restrictions over the connections
$A$ and $B$, in the form $\mathcal{L}_{\xi_h} \Gamma^{\lambda}_{\;\mu\nu}=0 $, we get the following possibilities for the functions $\gamma
_{1}(t)$, $\gamma_{2}(t)$:

\begin{itemize}
\item Case 1: For $\alpha=0$ we obtain two connections:
\begin{itemize}
 \item Connection $\Gamma_{1}$, derived from connection $A$ with $\gamma_{1}=\frac{\kappa
_{1}}{t}$, $\gamma_{2}=\frac{\kappa_{2}}{t}$
\begin{equation}  \label{connectionAg1}
\begin{split}
&  \Gamma_{\;tt}^{t}=\frac{\kappa_{2}}{t},\quad\Gamma_{\;tt}^{r}=\frac
{\kappa_{1}(\kappa_{1}-\kappa_{2}-1)}{c_{1}t^{2}},\quad\Gamma_{\;tr}%
^{r}=\Gamma_{\;t\theta}^{\theta}=\Gamma_{\;t\phi}^{\phi}=\frac{\kappa_{1}}%
{t},\\
&\Gamma_{\;\theta\theta}^{r}=-\frac{k}{c_{1}},\quad \Gamma_{\;rr}^{r}=\Gamma_{\;r\theta}^{\theta}=\Gamma_{\;r\phi}^{\phi}%
=c_{1},\quad\Gamma_{\;\phi\phi}^{r}=-\frac{k }{c_{1}}\Sigma_k(\theta)^2 ,\\
& \Gamma_{\;\phi\phi}^{\theta}=-\Sigma_k(\theta)\Sigma^{\prime}_k(\theta),\quad\Gamma_{\;\theta\phi
}^{\phi}=\frac{\Sigma^{\prime}_k(\theta)}{\Sigma_k(\theta)}.
\end{split}
\end{equation}
The corresponding nonmetricity scalar is obtained by direct substitution of $\gamma_1$, $\gamma_2$ and $\alpha=0$ in \eqref{nonmscA}
\begin{equation}
  Q = \frac{2 \left(3 \beta ^2 (\kappa_1-1)+k\right)}{\beta ^2 t^2}.
\end{equation}

\item Connection $\Gamma_{2}$, obtained from $B$, with $\gamma_{1}=\kappa_{1}$ and
$\gamma_{2}=\kappa_{2}t$
\begin{equation} \label{connectionBg2}
\begin{split}
&  \Gamma_{\;tt}^{t}=\frac{c_{1}\kappa_{1}(c_{2}\kappa_{1}-2)-\kappa_{2}%
-k}{\kappa_{2}t},\quad\Gamma_{\;tr}^{t}=c_{1}\left(  1-c_{2}\kappa_{1}\right)
,\quad\Gamma_{\;rr}^{t}=c_{1}c_{2}\kappa_{2}t,\\
& \Gamma_{\;\theta\theta}^{t}=\kappa_{2}t,\quad \Gamma_{\;\phi\phi}^{t}=\kappa_{2}t \Sigma_k(\theta)^2,\quad\Gamma_{\;tt}%
^{r}=\frac{\kappa_{1}(c_{1}\kappa_{1}+k)(c_{2}\kappa_{1}-1)}{\kappa_{2}%
^{2}t^{2}},\\
& \Gamma_{\;tr}^{r}=-\frac{c_{2}\kappa_{1}(c_{1}\kappa_{1}%
+k)}{\kappa_{2}t},\quad \Gamma_{\;rr}^{r}=c_{1}+c_{2} k +c_{1}c_{2}\kappa_{1},\quad\Gamma
_{\;\theta\theta}^{r}=\kappa_{1}, \\
& \Gamma_{\;\phi\phi}^{r}=\kappa_{1}%
\Sigma_k(\theta)^2,\quad\Gamma_{\;t\theta}^{\theta}=\Gamma_{\;t\phi}^{\phi}%
=-\frac{k+c_{1}\kappa_{1}}{\kappa_{2}t},\\
&  \Gamma_{\;r\theta}^{\theta}=\Gamma_{\;r\phi}^{\phi}=c_{1},\quad
\Gamma_{\;\phi\phi}^{\theta}=-\Sigma_k(\theta)\Sigma^{\prime}_k(\theta) ,\quad\Gamma_{\;\theta\phi
}^{\phi}=\frac{\Sigma^{\prime}_k(\theta)}{\Sigma_k(\theta)}.
\end{split}
\end{equation}
This time we obtain $Q$ by substitution in \eqref{nonmscB}
\begin{equation}
  Q = -\frac{2 \left(\kappa_2 \left(3 \beta ^2-2 \kappa_2\right)+\beta ^2 c_1 \left(c_2 \kappa_1^2-c_2 \kappa_2^2+2 \kappa_1\right)+\beta ^2 k (c_2 \kappa_1+2)-\kappa_2 k\right)}{\beta ^2 \kappa_2 t^2}.
\end{equation}

\end{itemize}

\item Case 2: For a generic $\alpha$, we get only one possible connection:

The connection $\Gamma_{3}$, given by $B$, with $\gamma_{1}=\kappa
_{1}t^{\alpha}$, $\gamma_{2}=\kappa_{2}t$ and $c_{1}=c_{2}=0$
\begin{equation} \label{connectionBg3}
\begin{split}
&  \Gamma_{\;tt}^{t}=-\frac{\kappa_{2}+k}{\kappa_{2}t},\quad\Gamma
_{\;\theta\theta}^{t}=\kappa_{2}t,\quad\Gamma_{\;\phi\phi}^{t}=\kappa_{2}%
t \Sigma_k(\theta)^2,\quad\Gamma_{\;tt}^{r}=-\frac{\kappa_{1}(\alpha\kappa
_{2}+k)t^{\alpha-2}}{\kappa_{2}^{2}},\\
&  \Gamma_{\;\theta\theta}^{r}=\kappa_{1}t^{\alpha},\quad\Gamma_{\;\phi\phi
}^{r}=\kappa_{1}t^{\alpha} \Sigma_k(\theta)^2,\quad\Gamma_{\;t\theta}^{\theta
}=\Gamma_{\;t\phi}^{\phi}=-\frac{k}{\kappa_{2}t},\\
&  \Gamma_{\;\phi\phi}^{\theta}=-\Sigma_k(\theta)\Sigma^{\prime}_k(\theta) ,\quad\Gamma
_{\;\theta\phi}^{\phi}=\frac{\Sigma^{\prime}_k(\theta)}{\Sigma_k(\theta)}.
\end{split}
\end{equation}
The relevant nonmetricity scalar is now
\begin{equation}
  Q = \frac{2 \left(\kappa_2 \left((2 \alpha -3) \beta ^2-(\alpha -2) \kappa_2\right)+k \left((\alpha -2) \beta ^2+\kappa_2\right)\right)}{\beta ^2 \kappa_2 t^2}.
\end{equation}

\end{itemize}

In what follows, for each of the above, self-similar  connections, we solve the corresponding field equations with the means to determine the functional form of the $f(Q)$ function in order for the former to admit the self-similar line-element (\ref{metricKS}) as a solution. We start from the vacuum case and subsequently we discuss the case where a perfect fluid matter source is included.

\section{Exact solutions} \label{sec4}

\subsection{Connection $\Gamma_{1}$}

For the first connection with nonzero components given by \eqref{connectionAg1}, we consider the spacetime (\ref{metricKS}) with $\alpha=0$. The
$t-r$ component of the equations of motion of the metric is equivalent to the
condition
\begin{equation}
c_{1}\dot{Q}f^{\prime\prime}(Q)=0. \label{eqtr1}%
\end{equation}
Due to $c_{1}\neq0$, since it appears in denominators in $\Gamma_{1}$, this
connection admits solutions which either have $Q=$const. or $f(Q)$ a linear
function of $Q$. In either case, the dynamics and the solutions produced are
indistinguishable from General Relativistic solutions with or without a
cosmological constant (the only way to evade this restriction would be to consider a fluid whose energy momentum tensor would introduce a nondiagonal $t-r$ component). We thus proceed with the consideration of the second connection, since for $\Gamma_{1}$ we obtain GR dynamics.

\subsection{Connection $\Gamma_{2}$}

Once more we require $\alpha=0$. The $t-r$ component of the metric equations of
motion yields now
\begin{equation}
\left(  2c_{1}(c_{2}\kappa_{1}+1)+c_{2} k \right)  \dot{Q}f^{\prime\prime}(Q)=0.
\label{eqtr2}%
\end{equation}
In order to depart from the cases that follow the same behavior as GR,
$f(Q)\sim Q$ or $Q=$const., we demand that
\begin{equation}
c_{2}=-\frac{2c_{1}}{2c_{1}\kappa_{1}+k},
\end{equation}
where $2c_{1}\kappa_{1}+k\neq0$. Notice that, if $2c_{1}\kappa_{1}+k=0$, then,
from Eq. \eqref{eqtr2}, we inescapably return to the case where we need to choose either
$\dot{Q}=0$ or $f^{\prime\prime}(Q)=0$.

We assume the perfect fluid energy momentum tensor defined as
\begin{equation}
  T_{\mu\nu} = (\rho +p) u_\mu u_\nu + p g_{\mu\nu},
\end{equation}
where $\rho$ and $p$ are the energy density and the pressure of the fluid. The $u^\mu$ is the comoving four-velocity satisfying $u^\mu u_\mu=-1$. The mixed tensor has components $T_{\;\;\nu}^{\mu
}=\mathrm{diag}(-\rho(t),p(t),p(t),p(t))$ and the equations of motion for the
metric, emanating from \eqref{fieldm}, read
\begin{align}
&  2t\left( k \beta^{2}+\frac{\beta^{2}c_{1}^{2}\left(  \kappa_{1}^{2}%
-\kappa_{2}^{2}\right)  }{2c_{1}\kappa_{1}+k}+\kappa_{2}^{2}\right)  \dot
{Q}f^{\prime\prime}(Q)-\kappa_{2}\left(  \beta^{2}\left(  6-t^{2}Q\right)
+2 k \right)  f^{\prime}(Q)\nonumber\\
& -\beta^{2}\kappa_{2}t^{2}f(Q)    =-2\beta^{2}\kappa_{2}t^{2}\rho
,\label{feq1gamma211}\\
&  2t\left(  \beta^{2}(2\kappa_{2}+k)-\kappa_{2}^{2}
+\frac{\beta^{2}c_{1}^{2}\left(  \kappa_{1}^{2}-\kappa_{2}^{2}\right)
}{2c_{1}\kappa_{1}+k}\right)  \dot{Q}f^{\prime\prime}(Q)-\kappa_{2}\left(
\beta^{2}\left(  t^{2}Q-2\right)  -2 k \right)  f^{\prime}(Q)\nonumber\\
& +\beta^{2}\kappa_{2}t^{2}f(Q)    =-2\beta^{2}\kappa_{2}t^{2}%
p,\label{feq1gamma222}\\
&  2t\left(  \frac{c_{1}^{2}\left(  \kappa_{1}^{2}+\kappa_{2}^{2}\right)
}{2c_{1}\kappa_{1}+k}+2\kappa_{2}+k\right)  \dot{Q}f^{\prime\prime}%
(Q)+\kappa_{2}\left(  2-t^{2}Q\right)  f^{\prime}(Q)\nonumber\\
& +\kappa_{2}t^{2}f(Q)    =-2\kappa_{2}t^{2}p, \label{feq1gamma233}%
\end{align}
while the equation for the connection \eqref{feq2} turns into
\begin{equation}
\left[  \beta^{2}c_{1}^{2}\left(  \kappa_{1}^{2}-\kappa_{2}^{2}\right)
+\left(  \beta^{2} k +\kappa_{2}^{2}\right)  (2c_{1}\kappa_{1}+k)\right]  \left(
tf^{\prime\prime\prime}(Q)\dot{Q}^{2}+t\ddot{Q}f^{\prime\prime}(Q)+3\dot
{Q}f^{\prime\prime}(Q)\right)  =0. \label{feq2gamma2}%
\end{equation}

The nonmetricity scalar is of the form
\begin{equation}
Q=\frac{\mathcal{A}}{t^{2}}, \label{Qgamma2}%
\end{equation}
where
\begin{equation}
\mathcal{A}=2\left[  \frac{2\kappa_{2}+k}{\beta^{2}}-\frac{2 k}{\kappa_{2}%
}-3-\frac{2c_{1}^{2}\left(  \kappa_{1}^{2}+\kappa_{2}^{2}\right)  }{\kappa
_{2}(2c_{1}\kappa_{1}+k)}\right]  . \label{Agamma2}%
\end{equation}
We shall investigate the conditions of admitting spacetime
\eqref{metricKS}, with $\alpha=0$, both in the presence of matter and in the vacuum. As a first step, let us observe that the combination $\beta^2\eqref{feq1gamma233} - \eqref{feq1gamma222}$, together with the use of \eqref{Qgamma2}, in order to substitute $t$ with respect to $Q$, leads to a differential equation for the $f(Q)$, which reads
\begin{equation}\label{genericf(Q)}
  f^{\prime}(Q) + S_k(c_1,\kappa_1,\kappa_2) Q f^{\prime\prime}(Q) =0,
\end{equation}
where $S_k(c_1,\kappa_1,\kappa_2)$ is a combination of the constants seen in the arguments and of $k$. The solution is of the form
\begin{equation}\label{genericf}
  f(Q) = \begin{cases}
           C_1 Q^\omega + C_2, & \mbox{if } S_k \neq 1 \\
           C_1 \ln Q +C_2, & \mbox{if } S_k = 1,
         \end{cases}
\end{equation}
where $\omega = \frac{S_k-1}{S_k}$ and with $C_1$, $C_2$ being constants of integration. We thus see that there are only two possibilities, either a power law or a logarithmic dependence on $Q$. The exact form of the function, will depend on the restrictions among the constants provided by the rest of the equations. Let us begin by considering the vacuum case.

\subsubsection{Vacuum}

Here, we assume $p=\rho=0$. It is straightforward to see that, in vacuum, the logarithmic function form, $f(Q)=C_1 \ln Q +C_2$, leads to no solution. So, for the time being, we need to restrict our analysis to the power law expression $f(Q)=C_1 Q^\omega + C_2$. For this particular form of $f(Q)$ theory, the equation \eqref{feq2gamma2}, reduces to
\begin{equation} \label{feq2gamma2const}
  \left[  \beta^{2}c_{1}^{2}\left(  \kappa_{1}^{2}-\kappa_{2}^{2}\right) + \left(  \beta^{2} k +\kappa_{2}^{2}\right)  (2c_{1}\kappa_{1}+k)\right] \left(\omega-1 \right) \left(\omega -2 \right) =0
\end{equation}
and can be satisfied in various distinct ways.

First of all, we exclude $\omega=1$ to avoid GR dynamics. The choice of $\omega=2$, when used into the rest of the field equations \eqref{feq1gamma211}-\eqref{feq1gamma233}, leads to a pure quadratic function solution $f(Q)\propto Q^2$, ($C_2=0$), with
\begin{align}
c_{1}=  &  \mp\frac{\sqrt{2\kappa_{2}+k }}{4\beta^{2}}\left[  \left(  \beta
^{2}\left(  \frac{4 k}{\kappa_{2}}+6\right)  +2\kappa_{2}+k \right)^{\frac
{1}{2}}\pm\sqrt{6\beta^{2}+2\kappa_{2}+k}\right] \\
\kappa_{1}=  &  \pm\kappa_{2} \left[ \frac{  \beta^{2}\left(  \frac{4 k}%
{\kappa_{2}}+6\right) }{2\kappa_{2}+k}+1\right]^{\frac{1}{2}}.
\end{align}
Note that the overall signs in the above relations for $c_{1}$ and $\kappa
_{1}$ are correlated. For example, if you take $c_{1}%
=-...$, then you need to combine it with $\kappa_{1}=+...$. For the above
solution the nonmetricity scalar assumes the simple form
\begin{equation}
Q=\frac{4(2\kappa_{2}+k)}{\beta^{2}t^{2}}
\end{equation}
and the final theory possesses two free parameters, $\kappa_2$ and $\beta$.

The other possibility is to satisfy \eqref{feq2gamma2const} without fixing $\omega$. For example, we may choose
\begin{equation}
\kappa_{2}=\pm\frac{\beta(c_{1}\kappa_{1}+k)}{\sqrt{\beta^{2}c_{1}^{2}%
-(2c_{1}\kappa_{1}+k)}}. \label{kgamma2a}%
\end{equation}
Of course, in this case, we need to demand that $c_{1}\kappa_{1}+k\neq0$ and
$\beta^{2}c_{1}^{2}-(2c_{1}\kappa_{1}+k)\neq0$, where the first inequality is
imposed because $\kappa_{2}$ cannot be zero, since it appears in denominators
inside the connection's components. At the end of this section, we shall separately
see what happens when $c_{1}\kappa_{1}+k=0$ and $\beta^{2}c_{1}^{2}%
-(2c_{1}\kappa_{1}+k)=0$ as special cases.

For the moment, let us continue with the generic case for which Eq.
\eqref{kgamma2a} holds. Then, the constant $\mathcal{A}$ involved in the
nonmetricity scalar \eqref{Qgamma2} becomes
\begin{equation}
\mathcal{A}=2\left[  \frac{k}{\beta^{2}}-3 \mp 4 \sqrt{\beta^2 c_1^2-(2 c_1\kappa_1+k)} \frac{c_{1}\kappa_{1}+k}%
{\beta\left(2 c_1 \kappa_1+k\right)}  \right]  .\label{Agamma2a}%
\end{equation}
Equation \eqref{feq1gamma211} implies that $C_2=0$, while the power, $\omega$, of the $f(Q)$ theory is given by
\begin{equation}\label{powermuKS}
  \omega=\frac{\mathcal{A}\beta^{2}}{(\mathcal{A}-6)\beta^{2}-2k}.
\end{equation}
With the use of expressions \eqref{Agamma2a} and \eqref{powermuKS}, the remaining equations, \eqref{feq1gamma222} and \eqref{feq1gamma233}, reduce to a single algebraic relation among constants. The latter can be written in the form
\begin{equation}\label{alggamma21}
  \beta  \left((\mathcal{A}+6) \beta ^2+2 k\right) \left(\beta ^2 c_1^2-2 c_1 \kappa_1- k\right)^{\frac{1}{2}} \mp 4 \left(3 \beta ^2+ k\right) (c_1 \kappa_1+ k) = 0,
\end{equation}
which in principle, after substituting $\mathcal{A}$ from \eqref{Agamma2a}, can be solved explicitly either with respect to $\beta$ or to $\kappa_1$. We refrain from giving the resulting complicated expressions and we just present some test values, in order to demonstrate that solutions with acceptable behaviors can emerge (e.g. solutions having a real metric with a Lorentzian signature).

For the Kantowski-Sachs case, $k=+1$, consider as an example the values: $c_{1}=-3$ and $\kappa_{1}=1$; then,
equation \eqref{alggamma21} yields the real solutions, $\beta\simeq \mp 0.608$ and $\beta
\simeq \pm 1.027$. These two, lead to power law
functions with $\omega \simeq 4.581$ and $\omega \simeq 0.669$. For the Bianchi type III case, $k=-1$, take $c_1=10$ and $\kappa_1=1/2$. The real solutions obtained by equation \eqref{alggamma21} are $\beta\simeq \pm 0.401$ and $\beta\simeq \mp 0.485$ leading to power-law theories with $\omega=1.181$ and $\omega=0.842$ respectively. We thus establish the existence of valid
$f(Q)\sim Q^\omega$ theories, compatible with these cosmological spacetimes in vacuum, when the
connection $\Gamma_{2}$ is used.

Let us mention that if we allow $\beta$ to be imaginary, the resulting spacetime becomes static (in the $k=+1$ case additionally spherically symmetric)
and no longer describes a cosmological model. The signature of the metric changes from $(-,+,+,+)$ to $(-,+,-,-)$ (the ordering of the coordinates being $(t,r,\theta,\phi)$). Thus, the $r$ variable, which acquires a different sign from the other three, becomes effectively the time variable, while $t$ is now spatial. A specific example of this in the $k=+1$ case
is given if we set $c_{1}=\kappa_{2}=0$. Then, Eq. \eqref{alggamma21} is satisfied by the
values $\beta=\pm\mathrm{i}(1\pm\sqrt{5})/2$. These values correspond
to a theory with $f(Q)$ function: $f(Q)\sim Q^{\frac{1}{4}(5\mp\sqrt{5})}$. We can write the line element in this case as
\begin{equation}
  ds^2 =  t^2 dr^2 - dt^2- \frac{\left(1\pm\sqrt{5}\right)^2}{4}  t^2 \left(d\theta^2 + \sin^2\theta d\phi^2 \right).
\end{equation}
In the above, $r$ is the time variable, while $t$ is the radial distance. It is naked singularity space, whose Ricci scalar diverges at the origin of the radius $t=0$.

At this point, it is interesting to see what would happen if we considered the
values which we excluded because of adopting the validity of Eq. \eqref{kgamma2a}. That is, the special
cases: $c_{1}\kappa_{1}+k=0$ and $\beta^{2}c_{1}^{2}-(2c_{1}\kappa_{1}+k)=0$.
By proceeding in a similar manner as in the generic case, if the first
equality holds, we have $c_{1}=-k/\kappa_{1}$. Then, we can satisfy
all the field equations under the conditions:
\begin{equation}
\kappa_{1}=\pm\sqrt{-k}\beta, \quad \kappa_2= \frac{\beta^2 k}{k+3\beta^2} \quad \text{and} \quad f(Q)\propto Q^{\frac{1}{2}\left( 1- \frac{k}{3 \beta^2}\right)}.
\end{equation}
On the other hand, if
$\beta^{2}c_{1}^{2}-(2c_{1}\kappa_{1}+1)=0$, then, we need to set
\begin{equation}
c_{1}=\pm\frac{\sqrt{-k}}{\beta} \quad \text{and again} \quad \kappa_2= \frac{\beta^2 k}{k+3\beta^2}, \quad f(Q)\propto Q^{\frac{1}{2}\left( 1- \frac{k}{3 \beta^2}\right)}.
\end{equation}
The aforementioned two cases are rather peculiar when considering $k=+1$, because, even though
the connection contains complex components, the metric is still Lorentzian and real. The same is true for the theory ($f(Q)$ function) and the nonmetricity scalar
($Q=\frac{2 k-6\beta^{2}}{\beta^{2}t^{2}}$ in both cases), which remain real. What is more, if we consider that the geodesic motion is given by minimizing the $ds^2$, which implies the appearance of the Christoffel symbols in the corresponding equations, then these imaginary units inside the connection would not appear even at that level. Thus, raising the question on whether there is some way to encounter these imaginary numbers in physical observables or if these solutions are to be considered as viable. A possibility of observing effects owed to non-metricity, would be by considering fermionic degrees of freedom. However, it has been noticed \cite{Delhom}, that in the case of a minimally coupled action, i.e. the typical action which in the case of Riemannian geometry leads to the Dirac equation, only the torsion would produce couplings to the $1/2$-spin field in the relevant equations of motion. Thus, in order to produce measurable effects which can be distinctly owed to nonmetricity, one may need to introduce more generalized actions for the fermions including nonminimal couplings. Still, in that case, we cannot comment for sure if the connection being complex would result in some particular effect or if all physical observables would in that case turn up to be real quantities, as it happens for example with the scalar $Q$, which is real even though $Q_{\lambda\mu\nu}$ is complex.

In order to summarize, we have seen that for $\Gamma_{2}$ the vacuum case accepts the cosmological
spacetime \eqref{metricKS}, for $\alpha=0$ as a solution, if the theory is
$f(Q)\propto Q^{\omega}$ or $f(Q)\propto Q^{2}$ depending on which of the
factors of \eqref{feq2gamma2const} becomes zero. Let's proceed to see what happens
with the inclusion of a perfect fluid.

\subsubsection{Perfect fluid matter content}

The equation for the connection, see Eq.
\eqref{feq2gamma2}, remains unaffected by the matter content. We can start by considering the power law case, $f(Q)=C_1 Q^\omega+C_2$, which again leads us to Eq. \eqref{feq2gamma2const}. As previously, the most generic solution is obtained by demanding that relation \eqref{kgamma2a} holds,
excluding once more the values $c_{1}\kappa_{1}+k=0$ and $\beta^{2}c_{1}%
^{2}-(2c_{1}\kappa_{1}+k)=0$, on which we are going to comment later.

We can easily eliminate the pressure by combining equations
\eqref{feq1gamma222} and \eqref{feq1gamma233} and, for the assumed $f(Q)$ function, we obtain a relation among constants which is satisfied if
\begin{equation}
\omega=1 \mp \frac{k(k+2c_{1}\kappa_{1})\sqrt{\beta^{2}c_{1}^{2}-(2c_{1}\kappa_{1}%
+k)}}{2\beta(k+c_{1}\kappa_{1})\left(  2\beta^{2}c_{1}^{2}+2c_{1}\kappa
_{1}+k\right)  }.
\end{equation}
We exclude the values for which $2\beta^{2}c_{1}^{2}+2c_{1}%
\kappa_{1}+k=0$ and $\omega=0$, as both of them lead to unacceptable $f(Q)$
functions for a physical theory (both imply $f(Q)=$constant).

By having made equations \eqref{feq1gamma222} and \eqref{feq1gamma233}
compatible with this choice of $f(Q)$, we can use either of them as a
definition for the pressure $p(t)$ and also the equation \eqref{feq1gamma211}
to give us the energy density $\rho(t)$. The result is
\begin{subequations}
\label{rhopfl1}%
\begin{align}
\rho &  =\frac{C_{1}\mathcal{A}^{\omega-1}\left(  \omega\left(
2k -(\mathcal{A}-6)\beta^{2}\right)  +\mathcal{A}\beta^{2}\right)  }{2\beta^{2}%
} t^{-2\omega} +\frac{C_{2}}{2}\\
p  &  =\frac{C_{1}\mathcal{A}^{\omega}}{2}\left[  \frac{\omega}{\mathcal{A}}\left(
\mathcal{A} - \frac{2}{\beta^{2}}\left(k \pm \frac{4\beta(\omega-1)(c_{1}\kappa
_{1}+k)}{\sqrt{\beta^{2}c_{1}^{2}-2c_{1}\kappa_{1}-k}}\right)
+8\omega-10\right)  -1\right]  t^{-2\omega}-\frac{C_{2}}{2}.
\end{align}

The additive appearance of $C_{2}$ is a trivial counterbalance of the
cosmological constant term in the $f(Q)$ function given by the first branch of \eqref{genericf}. So, we can simply eliminate it by setting $C_{2}=0$ in both
\eqref{genericf} and \eqref{rhopfl1}. The $\rho$ and $p$ of \eqref{rhopfl1},
with $C_{2}=0$, imply a linear barotropic equation of state of the form
$p=w\rho$, with
\end{subequations}
\begin{align}
w=\frac{p}{\rho}=  &  \frac{\mathcal{A}\beta^{2}}{\omega\left(  (\mathcal{A}-6)\beta
^{2}-2k \right)  -\mathcal{A}\beta^{2}}\times\nonumber\\
&  \left[  1-\frac{\omega}{\mathcal{A}}\left(  \mathcal{A}-\frac{2}{\beta^{2}%
}\left(k \pm \frac{4\beta(\omega-1)(c_{1}\kappa_{1}+k)}{\sqrt{\beta^{2}c_{1}%
^{2}-2c_{1}\kappa_{1}-k}}\right)  +8\omega-10\right)  \right],
\label{eqst1}%
\end{align}
where we see how the equation of state parameter, $w$, is calculated in terms of
the constants characterizing the model, $c_1$, $\kappa_1$ and $\beta$. The constant $\mathcal{A}$ is given as before from Eq. \eqref{Agamma2a}.

As we can see, the equation of state parameter involves enough independent constants to acquire any possible value associated with well-known matter contents. For example, if we consider the Bianchi type III case, $k=-1$, and for simplicity set $c_1=0$, then we obtain $w=-\frac{1}{3}\left(1\pm \frac{1}{\beta} \right)$ corresponding to an $f(Q)$ theory with power $\omega=1\mp\frac{1}{2\beta}$. This means that, for a radiation fluid $\beta=\mp \frac{1}{2}$, we would have an $f(Q)=Q^2$ theory. For different choices of the constants the resulting expressions are quite more involved, but in general there is enough freedom to associate any matter content to an appropriate theory and connection.

Let us briefly comment on what happens in the special cases, where
\eqref{kgamma2a} cannot be used because, either $c_{1}\kappa_{1}+k=0$ or
$\beta^{2}c_{1}^{2}-(2c_{1}\kappa_{1}+k)=0$ holds. As mentioned in the vacuum
section, these cases lead respectively to $\kappa_{1}=\pm \sqrt{-k}\beta$ or $c_{1}=\pm\frac{\sqrt{-k}}{\beta}$ (for $k=+1$, we encounter again the peculiarity of the connection being complex but the gravitational observables being real). The nonmetricity scalar
is once more $Q=\frac{2k-6\beta^{2}}{\beta^{2}t^{2}}$ and if we proceed with a
similar manner as before we end up with an $f(Q)$ function $f(Q)= C_{1}Q^{1-\frac{k}{6\kappa_{2}}}+C_{2}$ for both cases. We may again eliminate $C_{2}$ as it appears in a trivial manner. The resulting energy density and pressure are
\begin{subequations} \label{prg2a}
\begin{align}
\rho=  &  \frac{C_{1}\beta^{\frac{k}{3\kappa_{2}}-2}\left(  \beta^{2}%
(3\kappa_{2}-k)+\kappa_{2} k \right)  }{\kappa_{2}\left(  2 k -6\beta^{2}\right)
^{\frac{k}{6\kappa_{2}}}}t^{\frac{k}{3\kappa_{2}}-2}\\
p=  &  -\frac{C_{1}(3\kappa_{2}+k)\beta^{\frac{k}{3\kappa_{2}}-2}\left(
\beta^{2}(3\kappa_{2}-k)+\kappa_{2} k \right)  }{9\kappa_{2}^{2}\left(
2 k -6\beta^{2}\right)  ^{\frac{k}{6\kappa_{2}}}}t^{\frac{k}{3\kappa_{2}}-2},
\end{align}
\end{subequations}
and the equation of state is $p = -\left(\frac{1}{3} + \frac{k}{9\kappa_2} \right)\rho$. Of course, these expressions are true as long as $\kappa_{2}\neq k/6$. As we are going to see next, the $\kappa_{2} = k/6$ case is covered by a different $f(Q)$ function.

Finally, before proceeding to study the logarithmic case, we notice that \eqref{feq2gamma2const} can also be directly satisfied if we simply set $\omega=2$. This case is easily seen to lead to a solution with
\begin{equation} \label{kappa2alterg2}
   \kappa_2 = -\frac{k (2 c_1 \kappa_1 +k)}{2 \left(2 c_1 \left(\beta ^2 c_1 + \kappa_1 \right)+k\right)},
\end{equation}
characterized by a radiation equation of state $p= \rho/3$, where
\begin{equation} \label{rhospecialcase}
  \begin{split}
    \rho = & C_1 \mathcal{A}  \Bigg[\frac{12 c_1^2 k}{2 c_1 \left(\beta ^2 c_1+\kappa_1\right)+k}-\mathcal{A}-\frac{8 \beta ^2 c_1^2 k}{(2 c_1 \kappa_1+k)^2}-\frac{4 \left(4 \beta ^2 c_1^2+k\right)}{2 c_1 \kappa_1+k} \\
    & -\frac{8( \beta ^2 c_1^2- c_1 \kappa_1)}{k}\Bigg] t^{-4}.
  \end{split}
\end{equation}
In the above relations we once more ignored the contribution from $C_2$, considering just $f(Q)=C_1 Q^2$.

We now proceed to study what happens by setting $f(Q)=C_1 \ln Q + C_2$ in the equation of motion for the connection, Eq. \eqref{feq2gamma2}. In vacuum we saw that this logarithmic branch of \eqref{genericf} does not lead to a solution. The situation however changes with the inclusion of a perfect fluid. Equation \eqref{feq2gamma2} can only be satisfied by an appropriate choice of the constants in the bracket. The most general case corresponds again to $\kappa_2$, being given by \eqref{kgamma2a}; excluding of course the special situations where $c_{1}\kappa_{1}+k=0$ or $\beta^{2}c_{1}^{2}-(2c_{1}\kappa_{1}+k)=0$. Then, the field equations are solved by the following combination
\begin{subequations} \label{logsolrp1}
\begin{align}
  \rho=  & \frac{C_1}{2}  \left[\frac{2}{\mathcal{A}} \left(\frac{k}{\beta ^2}+3\right) +\ln \left(\frac{\mathcal{A}}{t^2}\right)-1\right]+\frac{C_2}{2}\\
  p = & -\frac{C_1}{2}  \left[\ln \left(\frac{\mathcal{A}}{t^2}\right)-\frac{1}{ \mathcal{A} \beta ^2} \left((\mathcal{A}-10) \beta^2 \pm \frac{8 \beta  (c_1 \kappa_1+k)}{\sqrt{\beta ^2 c_1^2-2 c_1 \kappa_1-k}}-2k\right)\right]-\frac{C_2}{2}
\end{align}
\end{subequations}
as long as the parameters $c_1$, $\kappa_1$ and $\beta$ satisfy the subsequent algebraic relation
\begin{equation}
  2 \beta  (c_1 \kappa_1+k) \left(2 \beta ^2 c_1^2+2 c_1 \kappa_1+k\right) \mp k (2 c_1 \kappa_1+k) \left(\beta ^2 c_1^2-2 c_1 \kappa_1-k\right)^{\frac{1}{2}} = 0 .
\end{equation}
The contribution of the constant $C_2$ is once more trivial so we might as well eliminate it. From the expressions for $p$ and $\rho$ we can write the equation of state for the fluid which is
\begin{equation}
  p = -\rho - \frac{2 C_1}{\mathcal{A}}\left(1 \mp \frac{2 (c_1 \kappa_1+k)}{\beta  \sqrt{\beta ^2 c_1^2-2 c_1 \kappa_1-k}}\right).
\end{equation}

Finally, we are left to examine the special cases $c_{1}\kappa_{1}+k=0$ and $\beta^{2}c_{1}^{2}-(2c_{1}\kappa_{1}+k)=0$. The first, through Eq. \eqref{feq2gamma2}, implies  $\kappa_1= \pm \sqrt{-k} \beta$ and the second $c_1 =\pm \sqrt{-k} / \beta$. Both cases lead to $\kappa_2 =k/6$, which is the value that was excluded from the power law solution that led us to expressions \eqref{prg2a}. Now, for the logarithmic $f(Q)=C_1 \ln Q $ function (we ignore the constant $C_2$), we obtain
\begin{subequations} \label{logsolrp2}
\begin{align}
\rho=  &  C_{1}\left[  \frac{3\beta^{2}}{k-3\beta^{2}}+\frac{1}{2}\ln\left(
\frac{2 k -6\beta^{2}}{\beta^{2}t^{2}}\right)  \right] \\
p=  &  C_{1}\left[  \frac{k-12\beta^{2}}{3\left(  k-3\beta^{2}\right)  }%
-\frac{1}{2}\ln\left(  \frac{2 k -6\beta^{2}}{\beta^{2}t^{2}}\right)  \right]  .
\end{align}
\end{subequations}
which leads to the equation of state $p = -\rho+\frac{C_{1}}{3}$. The nonmetricity scalars in these two special cases is again $Q=\frac{2 k-6\beta^{2}}{\beta^{2}t^{2}}$.

A comment is in order here, in what regards theories of this logarithmic form $f(Q)\sim \ln Q$. As can be seen by our solutions above, the nonmetricity scalar $Q$ is a function proportional to $t^{-2}$. Thus, there is an obvious singular point, in what regards the divergence at infinity of this geometric scalar, at $t=0$. The asymptotic future $t\rightarrow + \infty \Rightarrow Q\rightarrow 0$, however seems to also be problematic, in the sense that the action of the theory, as well as the physical expressions for the matter content, i.e. the $\rho$ and $p$, diverge. Hence, we seem to be dealing here with two singularities, one at the origin and another at infinity.

Before proceeding, let us mention that solutions like \eqref{prg2a} have the property of the matter tensor being also self similar, i.e. satisfying a relation of the form
\begin{equation} \label{matselfs}
\mathcal{L}_{\xi_h}T_{\mu\nu}= 2\sigma T_{\mu\nu},
\end{equation}
where $\sigma$ is a constant. It can be easily seen that, if we have a barotropic equation of state $p=w \rho$, then the above condition is satisfied if $\rho = \rho_0 t^{2(\sigma-1)}$. Which is something which is true for solution \eqref{prg2a} or \eqref{rhospecialcase}. The same does not hold for the logarithmic case, where we get \eqref{logsolrp1} or \eqref{logsolrp2}, for which a relation of the form \eqref{matselfs} does not hold.

\subsection{Connection $\Gamma_{3}$}

As discussed previously, this connection admits an arbitrary parameter $\alpha
$, but it has $c_{1}=c_{2}=0$. As a result of the latter, there is no
off-diagonal component in the metric field equations. The three remaining metric
equations are
\begin{align}
-2t\left( k \beta^{2}+\kappa_{2}^{2}\right)  \dot{Q}f^{\prime\prime}%
(Q)+\kappa_{2} \left(  (6-4\alpha)\beta^{2}-\beta^{2}%
t^{2}Q+2 k\right) f^{\prime}(Q)  &  +\beta^{2}\kappa_{2}t^{2}f(Q)=\nonumber\\
&  2\beta^{2}\kappa_{2}t^{2}\rho\label{feq1gamma311}\\
2t\left(  \kappa_{2}^{2}-\beta^{2}(2\kappa_{2}+k)\right)  \dot{Q}%
f^{\prime\prime}(Q)+\kappa_{2}\left(  \beta^{2}t^{2}Q-2\left(  \beta
^{2}+k\right)  \right)  f^{\prime}(Q)  &  -\beta^{2}\kappa_{2}t^{2}%
f(Q)=\nonumber\\
&  2\beta^{2}\kappa_{2}t^{2}p\label{feq1gamma322}\\
2t((\alpha-2)\kappa_{2}-k)\dot{Q}f^{\prime\prime}(Q)+\kappa_{2}\left(
t^{2}Q-2(\alpha-1)^{2}\right)  f^{\prime}(Q)  &  -\kappa_{2}t^{2}%
f(Q)=\nonumber\\
&  2\kappa_{2}t^{2}p, \label{feq1gamma333}%
\end{align}
while the single equation for the connection reads
\begin{equation}
\left( k \beta^{2}+\kappa_{2}^{2}\right)  \left[  tf^{\prime\prime\prime
}(Q)\dot{Q}^{2}+\left(  t\ddot{Q}-(\alpha-3)\dot{Q}\right)  f^{\prime\prime
}(Q)\right]=0. \label{feq2gamma3}%
\end{equation}
The nonmetricity scalar is once more $Q= \mathcal{A}/t^2$, where now
\begin{equation}
\mathcal{A}=\frac{2}{\beta^{2}\kappa_{2}}\left[  (\alpha-2)\left( k \beta^{2}%
-\kappa_{2}^{2}\right)  +\kappa_{2}\left(  (2\alpha-3)\beta^{2}+k\right)
\right].  \label{Qvalgamma3}%
\end{equation}

As previously, if we take a linear combination of \eqref{feq1gamma322} and \eqref{feq1gamma333} so as to eliminate the pressure, namely $\beta^2\eqref{feq1gamma333} - \eqref{feq1gamma322}$, we arrive once more to an equation of the following form
\begin{equation}\label{genericf(Q)2}
  f^{\prime}(Q) + S_k(\alpha,\beta,\kappa_2) Q f^{\prime\prime}(Q) =0,
\end{equation}
where again we have the admissible solutions \eqref{genericf}. That is, either a power law or a logarithmic function.

\subsubsection{Vacuum}

Let us start with the vacuum case $p=\rho=0$. Once more the logarithmic case $f(Q)= C_1 \ln Q +C_2$ leads to an incompatibility. So, we restrict the vacuum analysis to the power law case $f(Q)=C_1 Q^\omega +C_2$. First, we notice that in order to
satisfy (\ref{feq2gamma3}) we can either set $\kappa_{2}=\pm\sqrt{-k}\beta$
or impose a restriction on the power $\omega$, i.e. $\omega=(4-\alpha)/2$. The last case does not lead to an acceptable solution. On the other hand, the condition $\kappa_{2}%
=\pm\sqrt{-k}\beta$, leads us to the following set of solutions:
\begin{equation} \label{firstsetg3vac}
 \kappa_{2}= \sqrt{-k}\beta, \quad \alpha = \frac{\beta  \left(3 \beta -\sqrt{-k}\right) \pm \left(\beta -\sqrt{-k}\right) \left(9 \beta ^2+4 \beta  \sqrt{-k}+4 k\right)^{\frac{1}{2}}}{2 \beta  \left(2 \beta -\sqrt{-k}\right)}
\end{equation}
and
\begin{equation} \label{secondsetg3vac}
 \kappa_{2}= -\sqrt{-k}\beta, \quad \alpha = \frac{\beta  \left(3 \beta + \sqrt{-k}\right) \pm \left(\beta + \sqrt{-k}\right) \left(9 \beta ^2 - 4 \beta  \sqrt{-k}+4 k\right)^{\frac{1}{2}}}{2 \beta  \left(2 \beta + \sqrt{-k}\right)}.
\end{equation}
In both cases you have $C_2=0$, while the power $\omega$ of the $f(Q)\sim Q^{\omega}$ theory is given by
\begin{equation} \label{powermug3vac}
  \omega = \frac{\mathcal{A} \beta ^2}{\beta ^2 (4 \alpha +\mathcal{A} -6)-2 k},
\end{equation}
where $\mathcal{A}$ is obtained from \eqref{Qvalgamma3} after substitution of the above expressions of $\kappa_2$ and $\alpha$.

It is interesting to note that, in the $k=+1$ case, the only way in order to have an acceptable $f(Q)\propto Q^{\omega}$ theory, is to consider $\beta$ to be imaginary. Thus, instead of Kantowski-Sachs, the space must be static and spherically symmetric. To see this, set $k=+1$, for the first set \eqref{firstsetg3vac}, then the resulting power $\omega$ from \eqref{powermug3vac} is
\begin{equation} \label{powermuexg3}
\omega=\frac{(\beta-\mathrm{i})\left(3\beta-\mathrm{i} \mp  \sqrt{9\beta^{2}+4\mathrm{i}\beta
+4}\right)  }{(2\beta-\mathrm{i})\left(  3\beta \mp \sqrt
{9\beta^{2}+4\mathrm{i}\beta+4}\right)  }.
\end{equation}
For an imaginary $\beta$, we can have $\omega \in \mathbb{R}$ as long as, specific
bounds are set on the magnitude of the imaginary part. For example, if we take the plus sign of \eqref{powermuexg3} and set
$\beta=\mathrm{i}\sigma$, then we need to take the restrictions $\sigma\leq-\left(  2\sqrt{10}+2\right)  /9$ or
$\sigma\geq\left(  2\sqrt{10}-2\right)  /9$. Then, the $\omega$ takes the following
values: $\omega\in\lbrack5/2-\sqrt{5/2},5/4]$ for $\sigma\leq-\left(  2\sqrt
{10}+2\right)  /9$, $\omega\in(5/2+\sqrt{5/2},+\infty)$ for $1/2>\sigma
\geq\left(  2\sqrt{10}-2\right)  /9$ and $\omega\in(-\infty,\frac{1}{2})$ when
$1/2<\sigma$. Of course we need to exclude the value $\sigma=1$ which results
in $\omega=0$.

On the other hand, the $k=-1$ case of the Bianchi type III spacetime requires a real $\beta$, which however still needs to satisfy certain bounds in order to have the square roots in \eqref{firstsetg3vac}, \eqref{secondsetg3vac} and subsequently in $\omega$, yielding real numbers as results.

\subsubsection{Perfect fluid}

In the presence of a perfect fluid source, we start our analysis by considering the power law case $f(Q)=C_1 Q^\omega +C_2$. The equation of motion for the connection \eqref{feq2gamma3} reduces to
\begin{equation} \label{feq2gamma3red}
  \omega  (\omega-1)  (k \beta^2 +\kappa_2^2)(2\omega+\alpha-4) =0 .
\end{equation}
After leaving aside the values $\omega=0$ and $\omega=1$; the first because it does not lead to a valid theory and the second due to being indistinguishable from GR, we proceed with the two remaining choices: that of either fixing the $\kappa_2$ or the $\omega$. Let us start with the first, where we assume $\kappa_2 = \pm \sqrt{-k} \beta$. The field equations \eqref{feq1gamma311}-\eqref{feq1gamma333} yield the following energy density and pressure
\begin{align}
 \rho & = \frac{C_1 \mathcal{A}^{\omega-1}}{2 \beta^2} \left[ 2 \omega  \left((3-2 \alpha ) \beta ^2+k\right)-\mathcal{A} \beta ^2 (\omega -1) \right] t^{-2\omega} + \frac{C_2}{2} \\
 p & = \frac{C_1 \mathcal{A}^{\omega-1}}{2 \beta^2} \left[ \mathcal{A} \beta ^2 (\omega -1)-2 \omega  \left(\beta  \left(\beta  (5-4 \omega )\pm 4 \sqrt{-k} (\omega -1)\right)+k\right) \right] t^{-2\omega} - \frac{C_2}{2} ,
\end{align}
with the power of the theory being given by
\begin{equation}
  \omega = \frac{k-\left(\alpha -4\right) \alpha  \beta ^2 \mp 2 \beta  \sqrt{-k}}{2 \beta  \left(\alpha  \beta \mp \sqrt{-k}\right)} .
\end{equation}
The nonmetricity scalar ends up to be
\begin{equation}
  Q = \frac{\mathcal{A}}{t^2} = \frac{2}{\beta^2 t^2} \left[ (2 \alpha -3) \beta ^2 \pm 2 (2-\alpha ) \beta  \sqrt{-k}+k\right] .
\end{equation}
If we ignore once more the trivial constant $C_2$, i.e. set $C_2=0$, the equation of state parameter $w=p/\rho$ can be easily calculated in terms of the remaining constants of the model
\begin{equation}
  w = \frac{p}{\rho} = \frac{\beta  k (1-2 \alpha ) \pm \sqrt{-k} \left((\alpha -2) \alpha  \beta ^2-k\right)-\alpha ^2 \beta ^3 }{(\alpha -3) \beta  \left(\alpha ^2 \beta ^2+k\right)} .
\end{equation}
Notice, how in the case $k=+1$, we need again to require $\beta$ to be imaginary, so that $w$ and $\omega$ remain real. Thus, in that case we have once more static, spherically symmetric solutions. On the other hand, cosmological solutions emerge for $k=-1$, which corresponds to the Bianchi type III case.

We return now back to \eqref{feq2gamma3red}, and we investigate the other possibility, where $\omega = 2- \alpha/2$. Obviously we need to have $a\neq 4$ otherwise the power $\omega$ would vanish. The remaining equations \eqref{feq1gamma311}-\eqref{feq1gamma333} are solved by the combination
\begin{equation}
  \rho = 2^{1-\frac{\alpha}{2}} C_1 (\alpha-3) (\alpha -1)^{4-\alpha } t^{\alpha-4}, \quad p = \frac{\rho}{3-\alpha} , \quad \kappa_2 = \frac{k}{\alpha-2}
\end{equation}
for a theory of the form $f(Q) = C_1 Q^{2- \frac{\alpha}{2}}$ (once again we omit the additive constant in $f(Q)$).

We are finally left to study the logarithmic case, $f(Q)= C_1 \ln Q +C_2$. The equation of motion of the connection, Eq. \eqref{feq2gamma3}, now reduces to
\begin{equation} \label{feq2gamma3redb}
  (k \beta^2+ \kappa_2^2) (\alpha-4) =0,
\end{equation}
where we see that it includes the case $\alpha=4$, which we had to exclude from the power-law theory above. But first, let us consider the case $\kappa_2 = \pm \sqrt{-k} \beta$. The equations \eqref{feq1gamma311}-\eqref{feq1gamma333} are satisfied under the subsequent conditions: The energy density is
\begin{equation}
  \rho = \frac{C_1}{2} \ln\left(\frac{\mathcal{A}}{t^2}\right) + \frac{C_1}{\mathcal{A} \beta ^2}\left(2 k-\beta ^2 (4 \alpha +\mathcal{A}-6)\right) + \frac{C_2}{2}
\end{equation}
while the pressure becomes
\begin{equation}
  p = -\frac{C_1}{2} \ln\left(\frac{\mathcal{A}}{t^2}\right) + \frac{C_1}{\mathcal{A} \beta ^2} \left((\mathcal{A}-10) \beta ^2 \pm 8 \beta  \sqrt{-k}-2 k\right) - \frac{C_2}{2} .
\end{equation}
For the positive sign, i.e. $\kappa_2 = \sqrt{-k} \beta$, we have the additional relations
\begin{align}
  \beta & = \frac{\sqrt{-k} \left(1 \pm \sqrt{-\alpha ^2+4 \alpha +1}\right) }{\alpha  (4-\alpha )} \\ \label{Aconspcaseg3}
  \mathcal{A} & = 2 (\alpha -3) \left(\alpha+3 \mp 2 \sqrt{1-(\alpha -4) \alpha }\right)
\end{align}
and for the negative sign, $\kappa_2 = - \sqrt{-k} \beta$, we obtain
\begin{equation}
  \beta = -\frac{\sqrt{-k} \left(1 \pm \sqrt{-\alpha ^2+4 \alpha +1}\right) }{\alpha  (4-\alpha )}
\end{equation}
with the same expression for $\mathcal{A}$ as given in \eqref{Aconspcaseg3}. Again, we may set $C_2=0$ since its appearance is trivial in the relations. The equation of state for the fluid is then simply
\begin{equation}
  p =- \rho - \frac{2 C_1}{\mathcal{A} \beta } \left[ (\alpha +1) \beta \mp 2 \sqrt{-k} \right].
\end{equation}

Finally, we separately explore what happens if instead of fixing $\kappa_2$, we set $\alpha=4$ in Eq. \eqref{feq2gamma3redb}. Then, the system of equations \eqref{feq1gamma311}-\eqref{feq1gamma333} results in the following expressions
\begin{equation}
  \rho = \frac{C_1}{2} \ln \left(\frac{18}{t^2} \right) - C_1, \quad p = - \frac{C_1}{2} \ln \left(\frac{18}{t^2} \right), \quad \kappa_2 = \frac{k}{2},
\end{equation}
where we have already eliminated the additive constant $C_2$. The resulting nonmetricity scalar is simply $Q = 18/t^2$ and the equation of state reads $p=-\rho-C_1$.

\subsection{Dropping self-similarity for the connection}

The strategy we followed in the previous section involved the enforcement of the self-similarity - beside the metric - also at the level of the connection through Eq. \eqref{Liedercon}. We need to stress however, that this is not a necessary condition, in order for the equations of motion to be compatible with a self-similar metric of the form of Eq. \eqref{metricKS}. The reason behind the adoption of $\mathcal{L}_{\xi_h} \Gamma^\lambda_{\phantom{\lambda}\mu\nu}=0$ is the level of the complication that the field equations exhibit when no such condition is imposed. In this section we drop the requirement of self-similarity for the connection and try to see what kind of theories we may obtain in some very specific cases, where we can empirically make some assumptions which simplify the equations.

From the nondiagonal component of the field equations for the metric, it is easily derived that the generic connection A, as seen in \eqref{connectionA}, cannot have a solution that does not lead to $Q=$const. or $f(Q)$ being linear. So, the corresponding dynamics would be indistinguishable from GR. If we turn our attention to connection B of \eqref{connectionB}, the corresponding component of the field equations leads to
\begin{equation}
\left(  2c_{1}(c_{2}\gamma_1(t)+1)+c_{2} k \right)  \dot{Q}f^{\prime\prime}(Q)=0.
\label{eqtrBgen}%
\end{equation}
So, in order to avoid $Q=$const. or $f(Q)$ linear, we need to either impose
\begin{equation} \label{g1fromgen}
  \gamma_1 = - \frac{1}{c_2} - \frac{k}{2 c_1}
\end{equation}
or $c_1=c_2=0$. Up to this point we are in agreement with what we have seen in the previous sections, where either $\gamma_1$ had to be constant ($\Gamma_2$ connection) or the $c_i$'s should be zero ($\Gamma_3$ connection).

The nonmetricity scalar $Q$ becomes
\begin{equation}\label{Q1g1constant}
  \begin{split}
    Q =&  \left(\frac{2}{\beta ^2 t^2}-\frac{(c_2 k-2 c_1)^2}{4 c_1 c_2 \gamma_2^2}+c_1 c_2 t^{2 (\alpha -1)}\right) \dot{\gamma}_2 -\frac{(\alpha -3) t (c_2 k-2 c_1)^2}{4 c_1 c_2 t^2 \gamma_2} \\
    &+\frac{\left(-2 \alpha +(\alpha +1) \beta ^2 c_1 c_2 t^{2 \alpha }+2\right)}{\beta ^2 t^3} \gamma_2  +\frac{2 \left((2 \alpha -3) \beta ^2+k\right)}{\beta ^2 t^2}
  \end{split}
\end{equation}
when \eqref{g1fromgen} holds and
\begin{equation}\label{Q2czero}
  Q = 2 \left(\frac{(2 \alpha -3) \beta ^2+k+\dot{\gamma}_2}{\beta ^2 t^2}+\frac{(\alpha -3) k}{t \gamma_2}+\frac{k \dot{\gamma}_2}{\gamma_2^2}-\frac{(\alpha -1) \gamma_2}{\beta ^2 t^3}\right)
\end{equation}
if $c_1=c_2=0$. In the first case, $\gamma_1$ is fixed to a constant value, while in the second, due to $c_1=c_2=0$ it becomes irrelevant. Our strategy, in any of these cases, would be to choose $\gamma_2$ to be such a function of time, so that the relation $Q=Q(t)$ is easily invertible in order to express $t$ as a function of $Q$, and then as we previously did, use the field equations to derive a differential equation for the $f(Q)$.

In particular we will try to have relations of the form $Q=Q_0 t^{2\sigma}$. By doing so, even though we drop the condition of self-similarity upon the connection, we impose it in a sense on the nonmetricity scalar, since for such type of functions $\mathcal{L}_{\xi_h}Q=2\sigma Q$.

In what follows, we briefly present the end result of some solutions obtained in this manner.

\subsubsection{Set 1: $\alpha=0$, $c_2=-\frac{2}{c_1 \beta^2}$, $\gamma_2= \lambda t^3$ }

In this case we assume \eqref{g1fromgen} and hence for the nonmetricity scalar we have the expression \eqref{Q1g1constant}. For the specific values $\alpha=0$, $c_2=-\frac{2}{c_1 \beta^2}$ and $\gamma_2= \lambda t^3$, the latter reduces to
\begin{equation}\label{Qset1}
  Q = \frac{2 \left(k-3 \beta ^2\right)}{\beta ^2 t^2} .
\end{equation}
The theory which solves the field equations and leads to a self-similar metric $g_{\mu\nu}$ is now of an exponential form and for $k-3 \beta ^2\neq0$ we obtain
\begin{equation}\label{fQset1}
  f(Q) = C_1 e^{\frac{\beta ^2 k Q}{12 \lambda  \left(3 \beta ^2-k\right)}},
\end{equation}
where for similar reasons as previously we ignore an additive constant in the $f(Q)$ function.

The corresponding matter content has energy density and pressure
\begin{subequations}
  \begin{align}
    \rho & = \frac{C_1  \left(18 \beta ^4 \lambda +2 k^2 \lambda -k \left(12 \beta ^2 \lambda +\beta ^4 Q\right)\right)}{4 \lambda  \left(k-3 \beta ^2\right)^2} e^{\frac{\beta ^2 k Q}{12\lambda (3 \beta ^2 -k ) }} \\
    p & = \frac{C_1  \left(\beta ^6 k^2 Q^2-36 \lambda ^2 \left(k-3 \beta ^2\right)^3-2 \beta ^2 k \lambda  Q \left(k-6 \beta ^2\right) \left(k-3 \beta ^2\right)\right)}{72 \lambda ^2 \left(k-3 \beta ^2\right)^3} e^{\frac{\beta ^2 k Q}{12 \lambda (3\beta ^2 - k ) }}
  \end{align}
\end{subequations}
respectively. The resulting equation of state is rather complicated
\begin{equation} \label{eqstset1}
  p(\rho) = \frac{1}{9} \rho  \left(-\frac{k}{\beta ^2} + \frac{3 \beta ^2-k}{\beta ^2 W(\zeta \rho)}+6 W(\zeta \rho)\right)
\end{equation}
where $\zeta$ is the combinations of constants
\begin{equation}
  \zeta =\frac{ e^{\frac{1}{6} \left(\frac{k}{\beta ^2}-3\right)} \left(k-3 \beta ^2\right)}{3 \beta ^2 C_1}
\end{equation}
and $W(x)$ is the Lambert function, defined as the principal branch of the solution of the equation $W e^W = x$. Thus, the equation of state consists of a typical linear part, plus some additional terms given with a help of a non-elementary function.

An interesting observation regarding this result is that in the asymptotic future, $t\rightarrow +\infty$, we have $p=-\rho=$const., and the matter contribution is that of a cosmological constant, while the $f(Q)$ theory approximately becomes $f(Q)\sim Q +$const., implying GR dynamics. On the other hand, if we take $t\rightarrow 0$ and impose $\lambda k >0$, then $\rho\rightarrow 0$ and the expansion of \eqref{eqstset1} around zero yields
\begin{equation}
  p \simeq \frac{1}{9}  \left(3-\frac{2 k}{\beta ^2}\right)  \rho -\frac{C_1 }{3} e^{\frac{1}{2}-\frac{k}{6 \beta ^2}},
\end{equation}
which approximates a linear barotropic equation of state, but an $f(Q)$ function which diverges rapidly from the linearized GR dynamics.

\subsubsection{Set 2: $\alpha=0$, $c_1=\frac{c_2 k}{2}$, $\gamma_2= \lambda t^{-1}$ }

In this case we are also using \eqref{g1fromgen} and \eqref{Q1g1constant}, with the assumptions now that $\alpha=0$, $c_1=\frac{c_2 k}{2}$ and $\gamma_2= \lambda t^{-1}$. These choices leads again the non metricity scalar of \eqref{Q1g1constant} to become the one we see in \eqref{Qset1}. The field equations then imply the additional condition $c_2= \pm \frac{2}{\beta \sqrt{-k}}$ and the compatible $f(Q)$ function is now
\begin{equation}\label{fQset2}
   f(Q) = C_1 \left[ Q e^{\frac{k \left(k-3 \beta ^2\right)}{3 \beta ^2 \lambda  Q}} -\frac{k \left(k-3 \beta ^2\right) \text{Ei}\left(\frac{k \left(k-3 \beta ^2\right)}{3 Q \beta ^2 \lambda }\right)}{3 \beta ^2 \lambda } \right],
\end{equation}
where $\text{Ei}(x)=-\int_{-x}^{+\infty}\frac{e^{-z}}{z} dz$ is the exponential integral function. The corresponding energy density and pressure are
\begin{subequations}
  \begin{equation}
    \rho = C_1 \left[ \frac{Q \left(3 \beta ^2+k\right) e^{\frac{k \left(k-3 \beta ^2\right)}{3 \beta ^2 \lambda  Q}}}{2 \left(k-3 \beta ^2\right)}-\frac{k \left(k-3 \beta ^2\right) \text{Ei}\left(\frac{k \left(k-3 \beta ^2\right)}{3 Q \beta ^2 \lambda }\right)}{6 \beta ^2 \lambda } \right]
  \end{equation}
  \begin{equation}
    \begin{split}
      p = & \frac{C_1}{6 \beta ^2 \lambda  \left(k-3 \beta ^2\right)} \Bigg[ k \left(k-3 \beta ^2\right)^2 \text{Ei}\left(\frac{k \left(k-3 \beta ^2\right)}{3 Q \beta ^2 \lambda }\right) \\
      & -\beta ^2 e^{\frac{k \left(k-3 \beta ^2\right)}{3 \beta ^2 \lambda  Q}} \left(4 k \left(k-3 \beta ^2\right)+\lambda  Q \left(3 \beta ^2+k\right)\right) \Bigg] ,
    \end{split}
  \end{equation}
\end{subequations}
and they lead to an equation of state which of course we cannot write explicitly as $p(\rho)$, due to the appearance of the exponential integral function.

The first term of $f(Q)$ function in relation \eqref{fQset2}, which is of the form $Q e^{\frac{Q_0}{Q}}$, is reminiscent of the theory first proposed in \cite{ww8} and which is in good accordance with observational data. It is interesting to note that the $f(Q)$ of \eqref{fQset2} approaches General Relativity at the limit $Q\rightarrow \infty$, or equivalently at $t\rightarrow 0$, since in that limit the dominant term in $f(Q)$ is linear in $Q$.

As can be seen from the very particular solutions we obtained in this section, as we leave behind the assumption that the homothetic vector should be imposed on the connection, we are led to highly non-trivial $f(Q)$ functions. Unfortunately however, the problem becomes severely more complicated to deal in a generic manner.

\section{Reconstructing general flat connections from the corresponding
homothetic}

\label{sec5}

When the connection is flat, it is known \cite{Eisenhart,Koi} that there are coordinate
transformations which transform it to being identically zero. This has the
consequence that any two flat connections can be connected by a coordinate
transformation. Therefore, we can try to reconstruct the general connections
$A$, $B$ from the corresponding homothetic vectors $\Gamma_{1}$, $\Gamma_{2}$,
$\Gamma_{3}$.

To this end, let us use the fact that any flat, symmetric connection is described by a vector, whose components we choose to denote with $v^{\mu}$, as
\begin{equation}
\Gamma^{\lambda}{}_{\mu\nu}=\left(  {\Omega}^{-1} \right)  ^{\lambda}{}_{\rho}
\Omega^{\rho}{}_{\mu,\nu} \label{common 1}%
\end{equation}
where $\Omega^{\alpha}{}_{\beta}=v^{\alpha}{}_{,\beta}$.

The strategy we adopt is to find first the corresponding $v^{\mu}$'s for the three
homothetic connections. Let us consider connection $\Gamma_{1}$; the components of the contravariant vector ${v}^{\mu}$, in this case, are (we use the ordering of the components as $x^\mu=\{t,r,\theta,\phi\}$)
\begin{equation} \label{vecg1}
v^{\mu} = \left\{t^{\kappa_1} e^{c_1 r}  \Sigma'(\theta ) ,\; t^{\kappa_2+1} , \; t^{\kappa_1} e^{c_1 r} \Sigma(\theta )  \sin (\phi ), \; t^{\kappa_1} e^{c_1 r} \Sigma(\theta )  \cos (\phi )\right\}.
\end{equation}
By using this vector in \eqref{common 1} it is easy to verify that we obtain the homothetic connection $\Gamma_1$, whose nonzero components are seen in \eqref{connectionAg1}. Interestingly enough, we can use \eqref{vecg1} to obtain intuitively the vector which gives rise to the general flat connection $A$ given in \eqref{connectionA}. To do this we just need to remember that the homothetic vector $\xi_h$ and the condition $\mathcal{L}_{\xi_h}\Gamma =0$ basically fixes the time dependence. We may thus invert the process by inserting arbitrary functions of time in place of each different temporal expression we see in \eqref{vecg1}, i.e. make the conversions $t^{\kappa_1} \mapsto m_1(t)$ and $t^{\kappa_2+1} \mapsto m_2(t)$. In this manner we may write the more general vector
\begin{equation} \label{vecA}
v^{\mu} = \left\{m_1(t) e^{c_1 r}  \Sigma'(\theta ) ,\; m_2(t) , \; m_1(t) e^{c_1 r} \Sigma(\theta )  \sin (\phi ), \; m_1(t) e^{c_1 r} \Sigma(\theta )  \cos (\phi )\right\}.
\end{equation}
It can be checked now that the vector \eqref{vecA}, when used in \eqref{common 1}, returns the general flat connection $A$ of \eqref{connectionA} under the reparametrizations
\begin{equation} \label{map1}
m_{1}(t) = \exp\left[  \int\gamma_1(t) dt \right]  , \quad m_{2}(t) = \int \left[\exp\left(\int  \gamma_2(t)dt \right) \right] dt.
\end{equation}
We have thus gone backwards and obtained the general connection from the corresponding homothetic.

The same can be repeated for the second homothetic connection, $\Gamma_{2}$, for which, the
corresponding contravariant vector has the components
\begin{equation} \label{vecg2a}
v^{\mu} = \left\{\frac{e^{r (c_1+c_2 k)}}{t^{\frac{\kappa_1 (c_1+c_2 k)}{\kappa_2}}} ,\; \frac{e^{c_1 r}}{t^{\frac{c_1 \kappa_1+k}{\kappa_2}}} \Sigma'(\theta )  , \; \frac{e^{c_1 r}}{t^{\frac{c_1 \kappa_1+k}{\kappa_2}}} \Sigma(\theta ) \sin (\phi )  , \; \frac{e^{c_1 r}}{t^{\frac{c_1 \kappa_1+k}{\kappa_2}}} \Sigma(\theta ) \cos (\phi ) \right\}.
\end{equation}
Once more, by noticing the different time dependence, we may introduce the change $t^{-\frac{\kappa_1 (c_1+c_2 k)}{\kappa_2}} \mapsto m_1(t)$ and $t^{-\frac{c_1 \kappa_1+k}{\kappa_2}}\mapsto m_2(t)$ to write the more general vector
\begin{equation} \label{vecBa}
v^{\mu} = \left\{m_1(t) e^{r (c_1+c_2 k)} ,\; m_2 (t) e^{c_1 r} \Sigma'(\theta )  , \; m_2(t) e^{c_1 r} \Sigma(\theta ) \sin (\phi )  , \; m_2(t) e^{c_1 r} \Sigma(\theta ) \cos (\phi ) \right\}.
\end{equation}
This last vector generates the connection $B$ of \eqref{connectionB} through the reparametrization
\begin{equation} \label{map2}
\begin{split}
m_{1}(t) =  \exp\left[-\int \frac{\gamma_1(t) (c_1+c_2 k)}{\gamma_2(t)} dt\right], \quad m_2(t) = \exp\left[ - \int \frac{c_1 \gamma_1(t)+k}{\gamma_2(t)} dt \right] .
\end{split}
\end{equation}
So, once more we follow the same procedure to go backwards and recover the generic
connection $B$ from the homothetic $\Gamma_{2}$. We may notice that the above mapping, of relation \eqref{map2}, has a ``blind spot'' when $c_1=c_2=0$ since then, $m_1$ becomes a constant and we loose the parametrization from two $t$-functions to two $t$-functions. This special case is recovered by using the homothetic connection $\Gamma_3$ of \eqref{connectionBg3}, which exactly corresponds to the situation where $c_1=c_2=0$.

The vector which gives connection $\Gamma_3$ is
\begin{equation} \label{vecg3}
v^{\mu} = \left\{t^{-\frac{k}{\kappa_2}} \Sigma'(\theta ) ,\; r-\frac{\kappa_1 }{\alpha  \kappa_2}t^{\alpha }  , \; t^{-\frac{k}{\kappa_2}} \Sigma(\theta ) \sin (\phi ) , \; t^{-\frac{k}{\kappa_2}} \Sigma(\theta ) \cos (\phi )  \right\}.
\end{equation}
By following the same logic as before, we merely need to replace $t^{-\frac{k}{\kappa_2}} \mapsto m_1(t)$ and $\frac{\kappa_1 }{\alpha  \kappa_2}t^{\alpha } \mapsto m_2(t)$ . Then, we obtain the vector
\begin{equation} \label{vecBb}
v^{\mu} = \left\{m_1(t) \Sigma'(\theta ) ,\; r-m_2(t)  , \; m_1(t) \Sigma(\theta ) \sin (\phi ) , \; m_1(t) \Sigma(\theta ) \cos (\phi )  \right\},
\end{equation}
which truly reproduces the general flat connection $B$, under the reparametrization
\begin{equation} \label{map3}
\begin{split}
m_{1}(t) =  \exp\left[-\int \frac{k}{\gamma_2(t)} dt\right], \quad m_2(t) = \int \frac{\gamma_1(t)}{\gamma_2(t)} dt .
\end{split}
\end{equation}
In the next section we put in use the above vectors to derive and investigate in a straightforward manner the transformations between the corresponding connections.

\section{Transformations between different connections}

\label{sec6}

In this section we seek the general coordinate transformation that connects
the pairs ($\Gamma_{1}$, $A$), ($\Gamma_{2}$, $B$) and ($\Gamma_{3}$, $B$). We will utilize the
existence of the contravariant vector $v^{\mu}$ in terms of which a flat
connection is given by Eq. \eqref{common 1}.

Since, in any case, there exist coordinates in which $\Gamma^{\lambda}{}%
_{\mu\nu}$ vanishes, there must be a specific coordinate transformation
connecting the aforementioned pairs. The transformation sought for can be
found by equating the components of the two vectors $v_{(\text{generic})}%
^{\mu}=v_{(\text{homothetic})}^{\mu}$, where $v_{(\text{generic})}$ stands for
the vector that produces the generic connections, $A$ or $B$, with the arbitrary time
functions and $v_{(\text{homothetic})}$ for the one yielding $\Gamma_{1}$,
$\Gamma_{2}$, or $\Gamma_{3}$.

In what follows we use coordinates $(\bar{t},\bar{r},\bar{\theta},\bar{\phi})$ for
those referring to the full generic connection and $(t,r,\theta,\phi)$ for the ones in
which the homothetic connection is given. The Jacobian of the transformation
is $J^{\mu}{}_{\nu}=\frac{\partial x^{\mu}}{\partial \bar{x}^{\nu}}$ and the
equation mapping the connections from one coordinate system to another is
\begin{equation}
\label{treq}J^{\mu}{}_{\rho}\bar{\Gamma}^{\rho}{}_{\nu\kappa}-\Gamma^{\mu}%
{}_{\alpha\beta}J^{\alpha}{}_{\nu}J^{\beta}{}_{\kappa} -J^{\mu}{}_{\nu,\kappa}
=0,
\end{equation}
where we use $\bar{\Gamma}$ for the generic connections ($A$ or $B$) and
$\Gamma$ for the homothetic ones ($\Gamma_{1}$, $\Gamma_{2}$, or $\Gamma_{3}$).

Let us first choose the pair, ($A$, $\Gamma_{1}$). We write the generic vector given by \eqref{vecA} in the $\bar{x}$ system, and we have
\begin{equation}%
\begin{split}
v_{(\text{generic})}^{\mu} = \left\{m_1(\bar{t}) e^{c_1 \bar{r}}  \Sigma'(\bar{\theta} ) ,\; m_2(\bar{t}) , \; m_1(\bar{t}) e^{c_1 \bar{r}} \Sigma(\bar{\theta} )  \sin (\bar{\phi} ), \; m_1(\bar{t}) e^{c_1 \bar{r}} \Sigma(\bar{\theta} )  \cos (\bar{\phi} )\right\} .
\end{split}
\label{common1f}%
\end{equation}
We then take the homothetic vector of \eqref{vecg1} in the $x$ coordinates
\begin{equation}
\begin{split}
v_{(\text{homothetic})}^{\mu}  = \left\{t^{\kappa_1} e^{c_1 r}  \Sigma'(\theta ) ,\; t^{\kappa_2+1} , \; t^{\kappa_1} e^{c_1 r} \Sigma(\theta )  \sin (\phi ), \; t^{\kappa_1} e^{c_1 r} \Sigma(\theta )  \cos (\phi )\right\} .
\end{split}
\end{equation}
The coordinate transformation $x\rightarrow \bar{x}$, leading us from $\Gamma_1$ to $A$, is obtained by considering the equations $v_{(\text{generic})}^{\mu} =
v_{(\text{homothetic})}^{\mu}$, i.e.
\begin{equation}
t = m_2(\bar{t})^{\frac{1}{\kappa_2+1}} , \quad r = \bar{r} +\frac{1}{c_1} \left[\ln \left(\frac{m_1(\bar{t})}{m_2(\bar{t})^{\frac{\kappa_1}{\kappa_2+1}}} \right)\right] , \quad \theta =\bar{\theta}, \quad \phi = \bar{\phi}  ,
\end{equation}
which, as expected, satisfies \eqref{treq} for the connections produced by
$v_{(\text{generic})}$ and $v_{(\text{homothetic})}$.

It is interesting to note that with the above transformation, the line element
\begin{equation}
ds^{2}=-dt^{2}+t^{2}\left[ dr^{2}+\beta^{2}\left(d\theta^{2}+\Sigma_k(\theta)^{2}d\phi^2\right) \right], \label{metricKSnoa}%
\end{equation}
which we considered in the $\Gamma_1$ case, assumes now the expression
\begin{equation}
\begin{split} \label{metricKSnondiag}
  ds^2 = & \left[\left(\frac{m_2(\bar{t})^{\frac{1}{\kappa_2+1}}}{c_1} \left(\frac{m_1^{\prime}(\bar{t})}{m_1(\bar{t})}-\frac{\kappa_1 m_2^{\prime}(\bar{t})}{\kappa_2 m_2(\bar{t})+m_2(\bar{t})} \right) \right)^2 - \left(\frac{m_2(\bar{t})^{-\frac{\kappa_2}{\kappa_2+1}} m_2^{\prime}(\bar{t})}{\kappa_2+1}\right)^2 \right] d\bar{t}^2 \\
  & + \frac{2}{c_1}m_2(\bar{t})^{\frac{2}{\kappa_2+1}} \left(\frac{m_1^{\prime}(\bar{t})}{m_1(\bar{t})}-\frac{\kappa_1 m_2^{\prime}(\bar{t})}{\kappa_2 m_2(\bar{t})+m_2(\bar{t})}\right) d\bar{t} d\bar{r}\\
  & + m_2(\bar{t})^{\frac{2}{\kappa_2+1}} \left( d\bar{r}^2 + \beta^2 \left( d\bar{\theta}^2 + \Sigma(\bar{\theta})^2 d\bar{\phi}^2\right) \right) .
\end{split}
\end{equation}
The study we did regarding connection $\Gamma_1$ for the line element \eqref{metricKSnoa} is thus equivalent to taking the generic flat connection $A$, with the severely more complicated metric implied by \eqref{metricKSnondiag}. The $m_1$, $m_2$ are, as we mentioned in the previous section, connected to the $\gamma_1$ and $\gamma_2$ of the connection $A$ through Eq. \eqref{map1} (setting of course $\bar{t}$ instead of $t$).

By proceeding in the exact same manner we may obtain the transformation from $\Gamma_2$ to $B$, by equating the generic vector \eqref{vecBa} (written in the $\bar{x}$ coordinates) to the vector \eqref{vecg2a}. In this case, the result is
\begin{equation}
t=m_2(\bar{t})^{-\frac{\kappa_2}{k}} m_1(\bar{t})^{\frac{c_1 \kappa_2}{(k c_1+c_2)}}, \quad r= \bar{r} + \ln \left(\frac{ m_1(\bar{t})^{\frac{c_1 \kappa_1+k}{c_1 k+c_2}}}{ m_2(\bar{t})^{\frac{\kappa_1}{k}}} \right)  , \quad \theta =\bar{\theta}, \quad \phi = \bar{\phi} .
\end{equation}
Finally, by repeating the same process for \eqref{vecBb} (in $\bar{x}$) and \eqref{vecg2a} (in $x$) we obtain the mapping $x\rightarrow \bar{x}$ from the homothetic connection $\Gamma_3$ to the general connection $B$,
\begin{equation}
t=m_1(\bar{t})^{-\frac{\kappa_2}{k}}, \quad r= \bar{r} + \frac{\kappa_1 }{\alpha  \kappa_2}m_1(\bar{t})^{-\frac{\alpha  \kappa_2}{k}}-m_2(\bar{t})  , \quad \theta =\bar{\theta}, \quad \phi = \bar{\phi} .
\end{equation}
For each of the two last cases, the way the functions $m_1$, $m_2$ are connected to the original $\gamma_1$ and $\gamma_2$ is given by \eqref{map2} and \eqref{map3} respectively.

We thus see that the simple fact that all symmetric, flat connections can be mapped to each other does not trivialize the importance of considering distinct connections for a given spacetime. In a theory like $f(Q)$ gravity it is the pair ($g_{\mu\nu}$,$\Gamma^\lambda_{\;\mu\nu}$) which is of importance. For example, when considering a given line element, like \eqref{metricKSnoa}, we obtain different results for connections $\Gamma_1$ and $\Gamma_2$. It is true that we can map one connection to the other, but the result of such a mapping would change the line element as well. As happened in the case of the transformation that led us to \eqref{metricKSnondiag}.

\subsection{The coincident gauge}

We can briefly discuss now, what are the transformations that connect us to the coincident gauge. In order to find such a transformation we need to work in a similar manner as before. We may notice that in order to get out of equation \eqref{common 1} a zero connection, we need to use a vector $v^\mu$ whose components are linear in the new coordinates, e.g. $v^\mu=v^\mu_{(\text{coincident})}=\{\bar{r},\bar{t},\bar{\theta},\bar{\phi}\}$.

Let us consider the connection $\Gamma_1$ which is generated by the vector $v_{(\text{homothetic})}$ of Eq. \eqref{vecg1} and set $v_{(\text{homothetic})}^{\mu}(x)=v^\mu_{(\text{coincident})}(\bar{x})$. Then, we easily obtain the transformation $x\rightarrow \bar{x}$
\begin{equation} \label{coincident1}
\begin{split}
  & t = \left(\bar{t}\right)^{\frac{1}{\kappa_2+1}}, \quad r = \frac{\ln \left(\bar{t}^{-\frac{\kappa_1}{\kappa_2+1}} \sqrt{k \left(\bar{\theta} ^2+\bar{\phi} ^2\right)+\bar{r}^2}\right)}{c_1}, \quad \theta = \Sigma_k^{-1}\left(\frac{\sqrt{\bar{\theta} ^2+\bar{\phi} ^2}}{\sqrt{k \left(\bar{\theta} ^2+\bar{\phi} ^2\right)+\bar{r}^2}}\right) \\
  & \phi = \text{arccos}\left(\frac{\bar{\phi} }{\sqrt{\bar{\theta} ^2+\bar{\phi} ^2}}\right),
\end{split}
\end{equation}
where $\Sigma_k^{-1}$ is the inverse of the function $\Sigma_k(\theta)$, e.g. in the case k=-1 where $\Sigma_k=\sinh$, we have $\Sigma_k^{-1}= \text{arcsinh}$. The transformation \eqref{coincident1} takes us from the connection $\Gamma_1$, whose nonzero components we see in Eq. \eqref{connectionAg1}, to the coincident gauge where all the components become zero, i.e. $\Gamma^{\lambda}_{\phantom{\lambda}\mu\nu}=0$. It is easy to calculate the corresponding metric, although the expressions are rather complicated. In what follows we write the line element for the $k=-1$ case
\begin{equation} \label{met3coin}
  \begin{split}
  ds^2 =& \frac{\frac{\kappa_1^2}{c_1^2}-\bar{t}^{-\frac{2 \kappa_2}{\kappa_2+1}}}{(\kappa_2+1)^2} d \bar{t}^2 + \frac{\kappa_1 \bar{r} \bar{t}}{c_1^2 (\kappa_2+1) \left(\bar{\theta} ^2-\bar{r}^2+\bar{\phi} ^2\right)} d\bar{t} d\bar{r} \\
  &- \frac{\kappa_1 \bar{\theta} \bar{t} }{c_1^2 (\kappa_2+1) \left(\bar{\theta} ^2-\bar{r}^2+\bar{\phi} ^2\right)} d\bar{t} d\bar{\theta} - \frac{\kappa_1 \bar{\phi} \bar{t} }{c_1^2 (\kappa_2+1) \left(\bar{\theta} ^2-\bar{r}^2+\bar{\phi} ^2\right)} d\bar{t} d\bar{\phi} \\
  & \frac{\bar{t}^2 \left(\beta ^2 c_1^2 \left(\bar{\theta} ^2+\bar{\phi} ^2\right)+\bar{r}^2\right)}{c_1^2 \left(\bar{\theta} ^2-\bar{r}^2+\bar{\phi} ^2\right)^2} d\bar{r}^2 -\frac{\bar{\theta}  \bar{t}^2 \left(\beta ^2 c_1^2 \bar{r}+\bar{r}\right)}{c_1^2 \left(\bar{\theta} ^2-\bar{r}^2+\bar{\phi} ^2\right)^2} d\bar{r}d\bar{\theta} \\
  & -\frac{\bar{\phi}  \bar{t}^2 \left(\beta ^2 c_1^2 \bar{r}+\bar{r}\right)}{c_1^2 \left(\bar{\theta} ^2-\bar{r}^2+\bar{\phi} ^2\right)^2} d\bar{r}d\bar{\phi} + \frac{\bar{t}^2 \left(\beta ^2 \bar{\phi} ^2 \sinh ^2(\bar{\theta} )+\frac{\bar{\theta} ^2 \left(\bar{\theta} ^2+ \bar{\phi} ^2\right) \left(\beta ^2 c_1^2 \bar{r}^2+ \bar{\theta} ^2+ \bar{\phi} ^2\right)}{c_1^2 \left( \bar{\theta} ^2-\bar{r}^2+ \bar{\phi} ^2\right)^2}\right)}{\left(\bar{\theta} ^2+ \bar{\phi} ^2\right)^2} d\bar{\theta}^2 \\
  & + \frac{\bar{\theta}  \bar{t}^2 \bar{\phi}  \left(\frac{\left(\bar{\theta} ^2+ \bar{\phi} ^2\right) \left(\beta ^2 c_1^2 \bar{r}^2+\bar{\theta} ^2+\bar{\phi} ^2\right)}{c_1^2 \left(\bar{\theta} ^2-\bar{r}^2+\bar{\phi} ^2\right)^2}-\beta ^2 \sinh ^2(\bar{\theta} )\right)}{\left(\bar{\theta} ^2+\bar{\phi} ^2\right)^2} d\bar{\theta} d\bar{\phi} \\
  & + \frac{\bar{t}^2 \left(\beta ^2 \bar{\theta} ^2 \sinh ^2(\bar{\theta} )+\frac{\bar{\phi} ^2 \left(\bar{\theta} ^2+ \bar{\phi} ^2\right) \left(\beta ^2 c_1^2 \bar{r}^2+ \bar{\theta} ^2+ \bar{\phi} ^2\right)}{c_1^2 \left(\bar{\theta} ^2- \bar{r}^2+ \bar{\phi} ^2\right)^2}\right)}{\left(\bar{\theta} ^2+ \bar{\phi} ^2\right)^2} d\bar{\phi}^2 .
  \end{split}
\end{equation}
This is the line element of the Bianchi type III in the coincident gauge, if our starting connection is $\Gamma_1$. Similarly, the corresponding transformations having as starting points connections $\Gamma_2$ and $\Gamma_3$ can also be derived, although we refrain from giving the explicit relations here, since they become even more tedious.

The main point we want to stress is that, the fact that there always exists a coordinate system where the connection becomes zero, does not mean that we can blindly assume a priori the coincident gauge for any metric ansatz we make. For example, we see how both in the Kantowski-Sachs and the Bianchi type III LRS case of metric \eqref{metric}, none of the connections sharing the same symmetries as the metric is zero in the relevant coordinate system. The coincident gauge is achieved in different coordinate systems, where the metric acquires a quite more complicated form having all its components non-zero, as seen in \eqref{met3coin}.

\section{Conclusions}

\label{sec7}

The importance of self-similar solutions lies in the fact that they are used to model gravitational collapse and that of being equilibrium points in a dynamical systems analysis. They can be approached by more complicated solutions in certain limits; near the singularity or at the asymptotic future. Therefore it is useful to distinguish which types of modified theories can give rise to such spacetimes. In this work, we derived exact self-similar solutions in symmetric teleparallel $f\left(
Q\right)  $-theory for the anisotropic Kantowski-Sachs and Bianchi III
geometries. We considered the special form of these two spacetimes with
specific functions for the scale factors, so that a homothetic vector field
exist. For the latter, we presented all the independent flat connections which
were used to construct the $f\left(  Q\right) $-theory.

The gravitational Lagrangian density, given by the function $f\left(  Q\right)  $, was
constrained by the field equations, both in the case of vacuum and when a
matter source is introduced. The results are
summarized in Table \ref{tab1}. It is of special interests that we were able
to construct Static and Spherically Symmetric spacetimes (SSS).

For the connection $A$, we found that self-similar solutions exhibiting non-GR dynamics
exist when a matter source with nondiagonal terms is introduced in the field
equations. For the connection $B$, self-similar solutions were derived for
power-law $f\left(  Q\right)  =Q^{\omega}$ and
logarithmic $f\left(  Q\right) =C\ln
Q$  functions, as they are presented in Table \ref{tab1}.

Additionally, we presented a construction method which relates the specific
connections used for the self-similar solutions to the generic flat
connection for these spacetimes. Finally, we derived the transformations connecting certain pairs of the aforementioned connections.

Such an analysis provides us with important information about the behaviour of the
symmetric teleparallel $f\left(  Q\right)  $-theory in the early cosmology. The FLRW universe is of maximal symmetry, regarding the three-dimensional spatial slices, since it possesses a six-dimensional group of isometries. The next step, if we want to consider less symmetry and introduce anisotropy, is to investigate spacetimes whose spatial slices admit a four-dimensional algebra of Killing vectors (since a five-dimensional one cannot exist in three-space). Such is the motivation of choosing the cases of the Kantowski-Sachs and the Bianchi type III LRS spacetimes considered here. We
demonstrated that anisotropic, self-similar solutions are supported by this modified theory of
gravity and the necessary conditions to support them.

In our main results we saw how self-similarity, when imposed on both the metric and the connection, leads to either logarithmic models or power law functions $f(Q)$. The first class of solutions is plagued by singularities both at $Q\rightarrow 0$ and at $Q\rightarrow +\infty$, since the $f(Q)\sim \ln Q$, on which also the matter content depends, diverges. On the other hand, the power law solutions, are more physically oriented since the power $\omega$ of $f(Q)\sim Q^\omega$ can be chosen so as to signify small deviations from General Relativity. In addition, these theories can be used as a rough approximations of theories describing inflationary scenarios of the form $f(Q)=Q + \sigma Q^{1+|\epsilon|}$, in the epoch where $Q>>1$. In the solutions we obtained we saw that $Q\sim t^{-2}$, hence the region $Q>>1$ corresponds to the beginning of the universe, which is compatible with considering these theories as inflationary alternatives.

Finally, we experienced that, when relaxing the self-similarity condition on the connection, the scope of the theories that can admit self-similar metrics amplifies severely. Although it is quite complicated to extract solutions in this case, we managed to present some very specific cases, where the equations are simplified and certain $f(Q)$ functions can be recovered in analytic form. Both solutions we obtained in this case approach General Relativity in certain limits. In a future work we plan to investigate the generic behaviour of these
theories as well as to investigate the isotropic limits.

%

\begin{table}[tbp] \centering
\caption{Self-Similar non-GR solutions for the KS ($k=+1$) and BIII (LRS) ($k=-1$) spacetimes.}%
\begin{tabular}
[c]{cccc}\hline\hline
\multicolumn{4}{c}{\textbf{Connection} $\mathbf{A~}\left\{  \mathbf{~}%
\gamma_{1}\left(  t\right)  ,~\gamma_{2}\left(  t\right)
,~c_{1}\right\}  $}\\\hline\hline
$%
\begin{array}
[c]{c}%
\alpha=0~:~\mathbf{\Gamma}_{1}\\
\gamma_{1}=\frac{\kappa_{1}}{t},~\gamma%
_{2}=\frac{\kappa_{2}}{t}%
\end{array}
$ & \multicolumn{3}{c}{{\small Solution can exist only for
matter with nonzero }$t-r${\small \, component}}\\\hline\hline
&  &  & \\
\multicolumn{4}{c}{\textbf{Connection} $\mathbf{B~}\left\{  \mathbf{~}%
\gamma_{1}\left(  t\right) ,~\gamma_{2}\left(  t\right)
,~c_{1},~c_{2}\right\}  $}\\\hline\hline
$%
\begin{array}
[c]{c}%
\alpha=0:~\mathbf{\Gamma}_{2}\\
\gamma_{1}=\kappa_{1},~\gamma_{2} =\kappa_{2} t\\
c_{2}=\frac{-2c_{1}}{2c_{1}\kappa_{1}+1}
\end{array}
$
 & $T_{\mu\nu}=0$ & $\kappa_{2}$ \text{\small from} \eqref{kgamma2a} &  %
$f(Q)=Q^\omega$\, \text{{\small (includes SSS (k=+1) solutions)}}\\
 & & $\kappa_{2}$ \text{\small not from} \eqref{kgamma2a} & $f(Q)=Q^2$
\\
& $T_{\mu\nu} \neq0$ &  $\kappa_{2}$ \text{\small from} \eqref{kgamma2a} &
$%
\begin{array}
[c]{c}%
f\left(  Q\right)
=Q^{\omega} ~~,\text{ }p=w\rho\\
f\left(  Q\right)
=\ln Q~,~p=-\rho+C%
\end{array}
$
\\
&  & $\kappa_{2}$ \text{\small not from } \eqref{kgamma2a} & $%
\begin{array}
[c]{l}%
\kappa_{2}\neq\frac{k}{6} ,~f\left(  Q\right)
=Q^{\omega} ~~,\text{ }p=w\rho\\
\kappa_{2}=\frac{k}{6} ,~f\left(  Q\right)
=C\ln Q~,~p=-\rho+\frac{C}{3}\\
\kappa_{2} \, \text{\small from }\eqref{kappa2alterg2} ,~f\left(  Q\right)
=C\ln Q~,~p=-\rho+\frac{C}{3}
\end{array}
$\\\hline
$%
\begin{array}
[c]{c}%
\alpha\neq0:~\mathbf{\Gamma}_{3}\\
\gamma_{1}=\kappa_{1} t^{\alpha} ,~\gamma_{2} =\kappa_{2}t\\
c_{1} =0,~c_{2} =0%
\end{array}
$ & $T_{\mu\nu} =0$ & $\kappa_{2}=\pm \sqrt{-k} \beta~,~$ & $f\left(
Q\right)  =Q^{\omega}$, \,  {\small (k=+1) $\Rightarrow$ SSS}\\
&  & $\kappa_{2}\neq \pm \sqrt{-k}\beta$ ,  & {\small  no solution}\\
& $T_{\mu\nu}\neq0$ & $\kappa_{2}=\pm \sqrt{-k} \beta~,~$ & $%
\begin{array}
[c]{c}%
f(Q)=Q^\omega, \, p=w\rho , \, \text{\small ($k=+1$) SSS} \\
f(Q)=\ln Q, \, p=-\rho +C
\end{array}
$\\
&  &$\kappa_{2}\neq\pm \sqrt{-k} \beta $  & $%
\begin{array}
[c]{l}%
\alpha\neq 4, \, f\left(  Q\right) =Q^{2- \frac{\alpha}{2}}, \, p=\frac{\rho}{3-\alpha}\\
\alpha= 4, \, f(Q)= C \ln Q , \, p =- \rho + C
\end{array}
$\\\hline\hline
\end{tabular}
\label{tab1}%
\end{table}%

\begin{acknowledgments}
A.P. was supported in part by the National Research Foundation of South Africa
(Grant Numbers 131604). AP thanks the support of Vicerrector\'{\i}a de
Investigaci\'{o}n y Desarrollo Tecnol\'{o}gico (Vridt) at Universidad
Cat\'{o}lica del Norte through N\'{u}cleo de Investigaci\'{o}n Geometr\'{\i}a
Diferencial y Aplicaciones, Resoluci\'{o}n Vridt No - 096/2022.
\end{acknowledgments}

\end{document}